%% file: siren.tex
\documentclass[a4paper]{cas-dc}
\input{packages.tex}

\input{commands.tex}

\begin{document}

\input{sections/00_preamble}

\input{sections/01_introduction}

\input{sections/02_overview}

\input{sections/03_architecture}

\input{sections/04_features}
\input{sections/05_weighting}

\input{sections/06_examples}

\input{sections/07_conclusion}

\input{sections/08_acknowledge}

\bibliographystyle{elsarticle-num}
\bibliography{siren}

\pagebreak
\appendix

\input{appendices/a_computational.tex}
\input{appendices/b_code_examples.tex}
\input{appendices/c_lepton_injector_validation.tex}

\end{document}

%% file: packages.tex
\usepackage{CJKutf8}
\usepackage{amsmath}
\usepackage{url}
\usepackage{listings}
\usepackage{subfig}
\usepackage[capitalise]{cleveref}
\usepackage{natbib} 
\setcitestyle{square,numbers,sort&compress}
\usepackage{tabularx} 
\usepackage{array}
\usepackage{fontawesome} 
\usepackage{lettrine} 
\usepackage{fancyhdr} 
\usepackage{poemscol} 
\usepackage[htt]{hyphenat} 
\usepackage{xspace}
\usepackage[scaled=0.85]{beramono}
\usepackage[edges]{forest}
\usepackage{multicol}
\usepackage{siunitx}
\usepackage{seqsplit}
\usetikzlibrary{arrows.meta}


%% file: commands.tex
\newcommand{\siren}{\texttt{SIREN}\xspace}
\newcommand{\prevleptoninjector}{\texttt{LeptonInjector}\xspace}
\newcommand{\darknews}{\texttt{DarkNews}\xspace}
\newcommand{\python}{\texttt{Python}\xspace}
\newcommand{\cpp}{\texttt{C++}\xspace}

\newcommand{\RNum}[1]{\uppercase\expandafter{\romannumeral #1\relax}}

\newcommand{\cname}[1]{\renewcommand{\seqinsert}{\ifmmode\allowbreak\else\-\fi}\texttt{\seqsplit{#1}}\renewcommand{\seqinsert}{\ifmmode\allowbreak\else\hspace{0pt plus 0.02em}\fi}}
\newcommand{\bcname}[1]{\renewcommand{\seqinsert}{\ifmmode\allowbreak\else\-\fi}\textbf{\texttt{\seqsplit{#1}}}\renewcommand{\seqinsert}{\ifmmode\allowbreak\else\hspace{0pt plus 0.02em}\fi}}

\newcommand{\minerva}{MINER$\nu$A\xspace}
\newcommand{\miniboone}{MiniBooNE\xspace}

\definecolor{codegreen}{rgb}{0,0.6,0}
\definecolor{codegray}{rgb}{0.5,0.5,0.5}
\definecolor{codepurple}{rgb}{0.58,0,0.82}
\definecolor{backcolour}{rgb}{0.95,0.95,0.92}

\DeclareSIUnit \parsec {pc}

\lstdefinestyle{mystyle}{
    backgroundcolor=\color{backcolour},   
    commentstyle=\color{codegray},
    keywordstyle=\color{codegreen},
    numberstyle=\tiny\color{codegray},
    stringstyle=\color{orange},
    basicstyle=\ttfamily\footnotesize,
    breakatwhitespace=false,         
    breaklines=true,                 
    captionpos=b,                    
    keepspaces=true,                 
    numbers=left,                    
    numbersep=5pt,                  
    showspaces=false,                
    showstringspaces=false,
    showtabs=false,                  
    tabsize=2
}

%% file: sections/00_preamble.tex
\let\WriteBookmarks\relax
\def\floatpagepagefraction{1}
\def\textpagefraction{.001}

\author[1,2]{Austin Schneider}[
    orcid=0000-0002-0895-3477
]
\cormark[1]
\ead{aschn@mit.edu}

\author[3]{Nicholas W. Kamp}[
    orcid=0000-0001-9232-259X
]
\cormark[1]
\ead{nkamp@fas.harvard.edu}

\author[3]{Alex Y. Wen}[
    orcid=0009-0009-4869-7867
]
\cormark[0]

\affiliation[1]{
    organization={Los Alamos National Laboratory},
    city={Los Alamos},
    citysep={}, 
    postcode={}, 
    state={NM},
    country={United States}
}

\affiliation[2]{
    organization={Massachusetts Institute of Technology},
    city={Cambridge},
    citysep={}, 
    postcode={}, 
    state={MA},
    country={United States}
}

\affiliation[3]{
    organization={Department of Physics \& Laboratory for Particle Physics and Cosmology, Harvard University},
    city={Cambridge},
    citysep={}, 
    postcode={02138}, 
    state={MA},
    country={United States}
}


\cortext[cor1]{Corresponding authors}

\shortauthors{A. Schneider \textit{et~al.}}

\title [mode = title]{\siren: An Open Source Neutrino Injection Toolkit {\raisebox{-2.50\depth}{{\href{https://github.com/Harvard-Neutrino/SIREN}{\huge\color{BlueViolet}\faGithub}}}}}

\shorttitle{\texttt{SIREN}}

\begin{abstract}
Modeling of rare neutrino processes often relies on either simple approximations or expensive detector simulations.
The former is often not sufficient for interactions with complex morphologies, while the latter is too time-intensive for phenomenological studies.
We present \siren~(Sampling and Injection for Rare EveNts), a new tool for neutrino phenomenology and experimental searches alike that enables accurate interaction and detector geometry modeling without the overhead of detailed detector response simulations.
\siren handles the injection of rare process final states and the associated weighting calculations with the speed needed for phenomenological investigations and the detail necessary for dedicated experimental searches.
The extensible design of \siren~allows it to support a wide range of experimental designs and Beyond-Standard-Model neutrino interactions.
Users need only specify the physical process, detector geometry, and initial neutrino flux under consideration before they can accurately simulate a model in their detector of choice.
We demonstrate the capability of \siren~through two examples: (1) Standard Model $\nu_\mu$ deep inelastic scattering in IceCube, DUNE, and ATLAS; and (2) heavy neutral lepton interactions in \miniboone, \minerva, CCM.
A variety of detector geometry descriptions, interaction cross sections, and neutrino fluxes are also provided for users to get started with immediately.
\end{abstract}


\maketitle

%% file: sections/01_introduction.tex
\section{Introduction}
\label{sec:intro}

The simulation of neutrino interactions can be divided into three steps: 1) the injection of neutrino interactions, 2) the simulation of the detector response, and 3) the weighting of the simulation results.
In phenomenological studies of the neutrino sector, it is common practice to use a series of simple approximations for these three steps, rather than employing expensive simulations.
However, for processes with multiple interaction vertices or otherwise large spatial extent, the details of the detector's geometry become important for correctly modeling sensitivity to such interactions.
Full-scale detector simulations like \cname{GEANT4}~\cite{GEANT4:2002zbu} can be used to correctly model the detector's response to such interactions, but such simulations have a large overhead and can be difficult to customize for each new interaction.
It is instead possible to account for the majority of these geometric effects in the injection and weighting steps, which have lower computational overhead.
This trade-off is possible because the low interaction cross section of the initial neutrino interactions means that these geometric effects do not need to be modeled with the same level of precision as their potentially visible products.
This approach is particularly advantageous for phenomenological studies of physics beyond the Standard Model (BSM) coupled to the neutrino sector, which gain additional modeling accuracy without the associated computational cost.
In experimental collaborations, the ability to re-weight simulation results to different physical hypotheses has become critical.
Detailed detector simulations in accelerator neutrino experiments and neutrino telescopes are often prohibitively expensive, and robust re-weighting techniques allow these simulations to be re-used more widely.
The separation of new physics modeling from the detector simulation and a robust approach to weighting thus can vastly reduce the effort needed to test new physics models for both phenomenologists and experimental collaborations.
Although we primarily discuss the problems that arise when simulating neutrino interactions, these same considerations apply to rare processes in a variety of BSM scenarios.

To address these issues we introduce \siren--a new software tool for the injection and weighting of neutrino interactions that centers speed, re-weightability, and extensibility for BSM interactions and complex detector geometries.
\siren gives users the ability to easily change detectors, interaction models, and injection strategies.
Additionally, \siren allows users to re-weight to a physical scenario regardless of what different detector geometries, interaction models, and injection strategies the underlying simulation sets were generated with, a process which is only limited by the overlap between physical and generation parameter space.
The extensible architecture of \siren allows users to easily add their own models, detector configurations, and injection strategies.
At the time of writing, a variety of these are already distributed as part of the \siren Python package.
Although the primary user interface is in Python, the core routines of \siren are written in \cpp to lower computation costs.
\siren's underlying architecture allows nearly every sub-component to be extended in the \cpp code-base, and a subset of these components can be extended by users through the Python interface.
Development of \siren grew out of efforts to improve the \prevleptoninjector event generator developed within the IceCube collaboration~\cite{IceCube:2020tcq}, and so shares design elements and some underlying code with \prevleptoninjector.

\siren allows the user to specify the distributions from which to sample the initial neutrino properties (energy, direction, and helicity) as well as any possible primary and secondary interactions, requiring only the total and differential cross section or decay width for each interaction.
Support exists within \siren for several neutrino interaction models, including neutrino-portal BSM scenarios via a custom interface with the \darknews software package~\cite{Abdullahi:2022cdw}.
The user can then separately specify the geometric configuration of the detector using configuration files that define the three-dimensional shape and atomic composition of each detector subcomponent.
The construction of new configuration files is straightforward, and at the time of writing \siren includes example configuration files for the following experiments: MiniBooNE~\cite{MiniBooNE:2008paa}, CCM~\cite{CCM:2021leg}, DUNE~\cite{DUNE:2020txw}, \minerva~\cite{MINERvA:2013zvz}, IceCube~\cite{IceCube:2016zyt}, ATLAS~\cite{ATLAS:2008xda}, and Hyper-K~\cite{Hyper-Kamiokande:2018ofw}.

\siren uses biased and physical distributions to sample properties of the primary neutrino, its interaction, and the interactions of subsequent secondary particles, if desired.
After generation, \siren corrects for the biased injection distributions across all particles to produce a physical weight.
The reweighting procedure of \siren can be used to weight between different interaction models, weight simulation to different detector models, and combine simulation sets that use different injection strategies.

In combination, the tools and capabilities provided by \siren make it easy for users to start simulating a new scenario, and afford users the ability to maximize the impact of more detailed simulation efforts through more targeted injection and the reuse of existing simulations.

Within this article, \cref{sec:overview} provides a broad outline of how \siren functions and the design choices made in the package.
We cover the architecture of \siren in \cref{sec:architecture}, including a detailed description of the methods used for event injection, interaction specification, and geometry configuration.
Next, \cref{sec:features} describes the important features of \siren, including the generalized model interface, extensible geometry setup, efficient event generation, and flexible injection methodology.
We discuss the calculation of physical event weights in \cref{sec:weighting}.
\Cref{sec:examples} provides a series of examples to demonstrate the capabilities of \siren, including $\nu_\mu$ deep inelastic scattering in IceCube, DUNE, and ATLAS as well as dipole-coupled heavy neutral lepton interactions in MiniBooNE, MINERvA, and CCM.
Finally, \cref{sec:conclusion} concludes with a summary of the unique features of \siren presented in this article.

%% file: sections/02_overview.tex
\section{Overview of \siren}
\label{sec:overview}

The most powerful features of \siren are

\begin{enumerate}
    \item its extensibility,
    \item its comprehensive injection scheme, and
    \item its reweighting capabilities.
\end{enumerate}

We define \cname{injection} as the sampling of particle properties from their corresponding distributions, and \cname{weighting} as the process of correcting for the differences between the physical distributions of particle properties and those used during injection.
In this view, we conceptually need only concern ourselves with a particle's properties and the probability distributions that govern them.
Generally speaking, these multi-dimensional probability distributions can be quite complex, but in the physical scenarios we are concerned with, they can be decomposed.
\siren approaches the problems of injection and weighting by breaking up the probability distribution of particle properties into a series of conditional probability distributions, and breaking up physical events into individual interactions.

Within \siren, each injection and weighting routine is primarily concerned with the \cname{InteractionRecord}, a structure that holds properties of the incident primary particle, target particle, and secondary particles.
This structure is referenced at every stage to assign and query the properties of the particles involved in the interaction.

In the physical scenarios we are ultimately modeling, the probability distributions of particle properties depend strongly on the interactions it can undergo, and the detector geometry.
Our injection methodologies often depend similarly on these properties to provide appropriate coverage of the parameter space.
For this reason, the injection and weighting routines of \siren are divided into three major components: injection distributions, interactions, and detector geometry calculations.
These components are coordinated by the \cname{Injector} class to perform the event injection and are queried by the \cname{Weighter} class to perform weighting operations.

In practical terms, the process of injecting an event can be thought of as sequentially passing the \cname{InteractionRecord} to each injection distribution where the corresponding properties of the interaction are sampled and recorded, such as its direction, energy, initial position, and interaction vertex.
This sequential approach naturally aligns with the idea of conditional probability distributions, as each injection distribution is provided with the properties assigned by previous distributions.
To enable injection distributions more closely aligned with the physical processes, injection distributions also receive the contextual information of the detector geometry and available interactions.
In this way, our conditional probability distributions depend on the detector geometry and available interactions in addition to previously assigned particle properties.

Once the properties of a particle have been assigned and an interaction vertex selected, a target and interaction type are randomly chosen in a manner proportional to the interaction rate of the available processes.
In \siren these are any combination of $2 \to n$ or $1 \to n$ processes, that in practice are a specialization of the \cname{CrossSection} or \cname{Decay} class, respectively.
These classes are then responsible for choosing the secondary particles and assigning the appropriate kinematics.

Some final states will depend on multiple interactions occurring, so \siren enables the same injection and weighting procedure for secondary particles and has a mechanism to control what secondaries are allowed to have subsequent interactions.
The main difference for secondary particle injection is that the previous interaction fixes the secondary's starting point and kinematic properties, so \siren only samples to the secondary interaction vertex through a new injection distribution.
A new \cname{CrossSection} or \cname{Decay} class then determines the kinematics of outgoing particles from this secondary interaction.
In this paradigm, the distribution of secondary energies and directions is assumed to be independent of the \cname{DetectorModel} for a given target.
A future update is planned that will relax this assumption, and allow biasing of the secondary kinematic distributions.
\siren permits an arbitrary number of secondary interactions during simulation, which each have their own \cname{InteractionRecord} and together populate the nodes of an \cname{InteractionTree}.

\begin{figure*}
    \centering
    \includegraphics[width=0.95\textwidth]{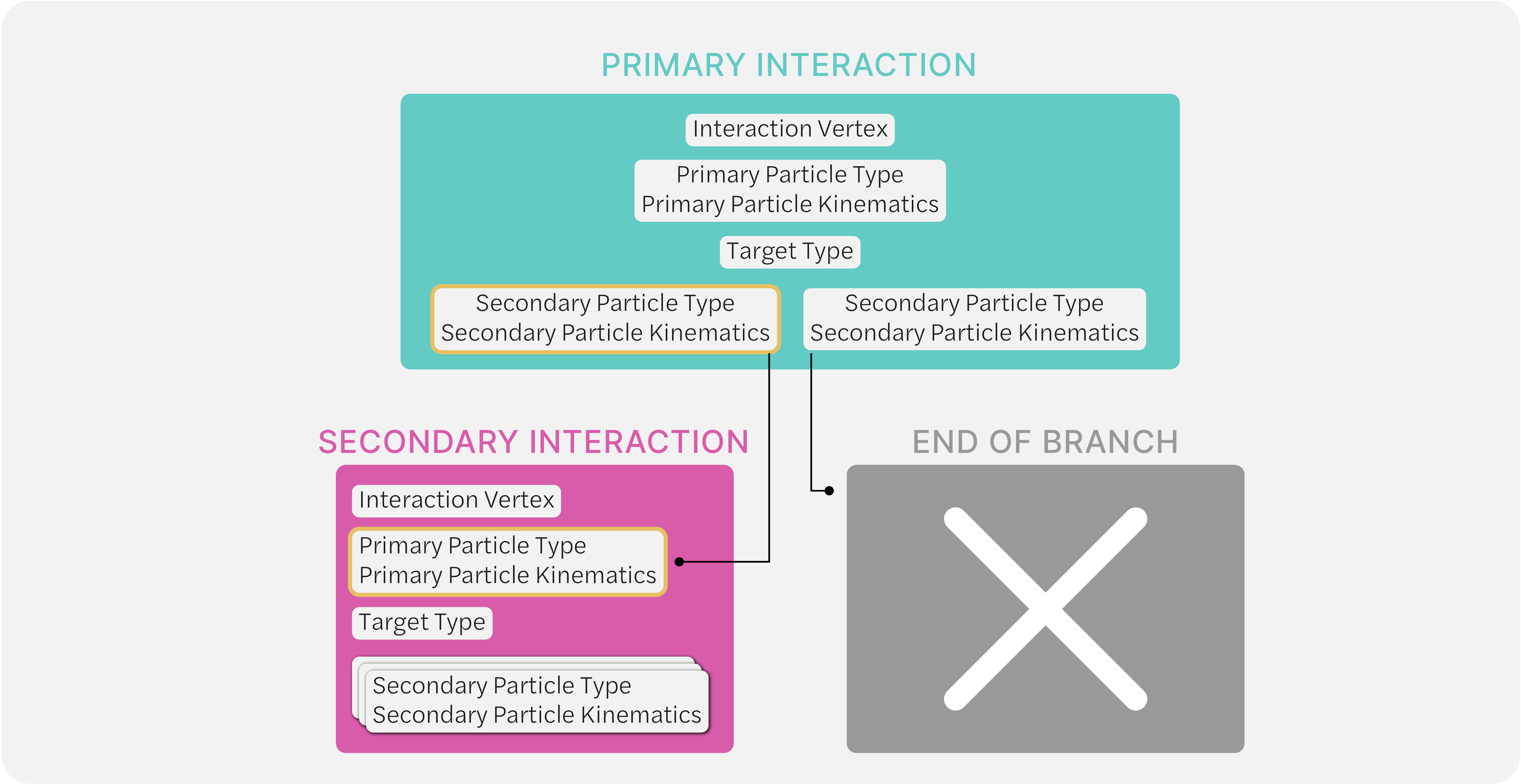}
    \includegraphics[width=0.95\textwidth]{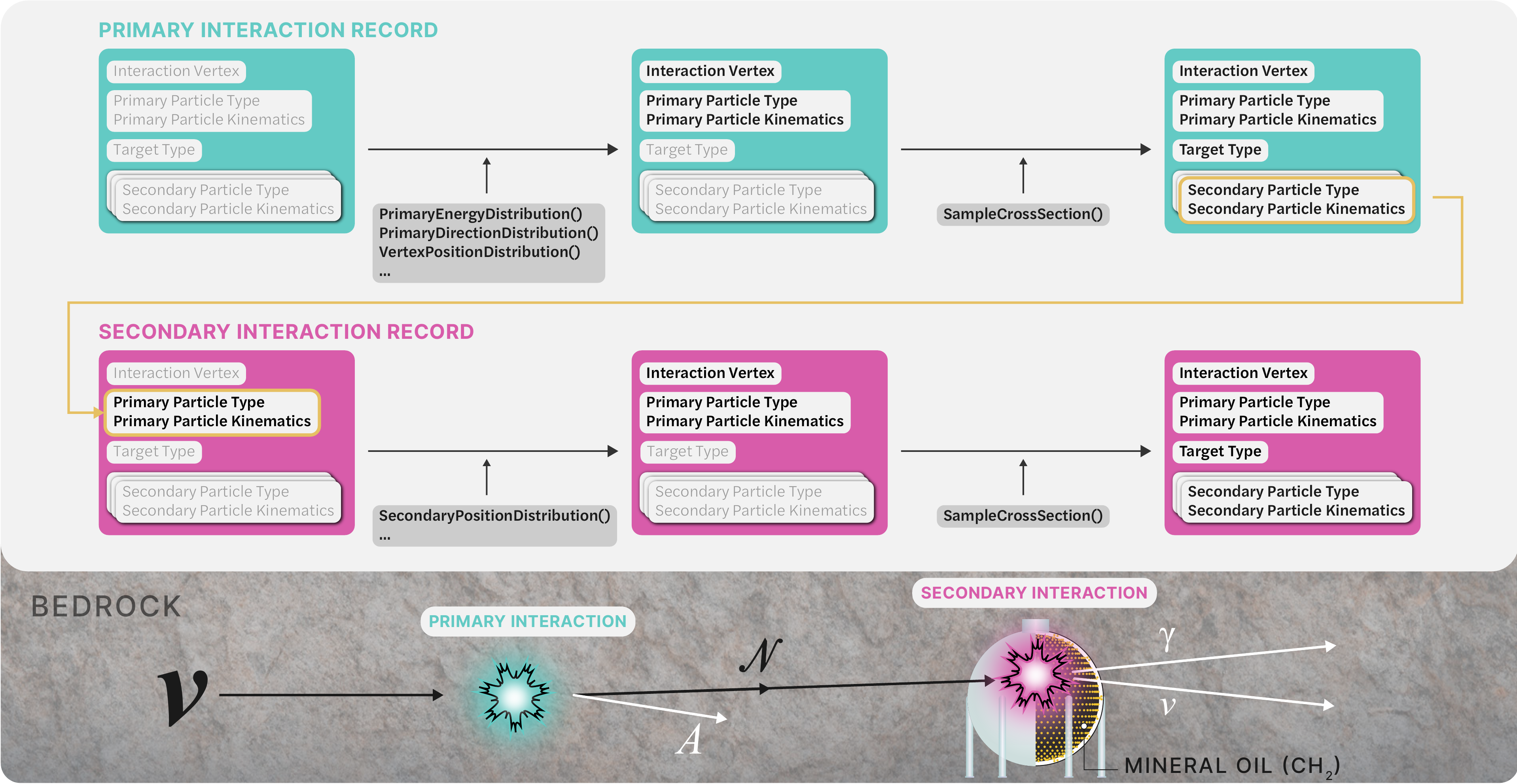}
    \caption{This figure depicts an example injection in \siren. We consider the production and decay of dipole-coupled HNLs in MiniBooNE. The top panel shows the \texttt{InteractionTree} object, which consists of a tree of \texttt{InteractionRecord} objects with a single primary interaction at the top. In this case, there is only one secondary interaction in the tree. The bottom panel shows the steps through which \siren fills the \texttt{InteractionRecord} for both interactions. Also depicted is the corresponding physical picture, in which an incoming neutrino interacts with a nucleus in the bedrock surrounding MiniBooNE to produce an HNL, which decays into a photon and neutrino inside the MiniBooNE detector. The white arrows refer to secondary particles that are no longer tracked as part of the \texttt{InteractionTree}.}
    \label{fig:siren_injection_diagram}
\end{figure*}

To outline the injection process, \Cref{fig:siren_injection_diagram} depicts the sequence of operations in \siren that would occur for the production and subsequent decay of a dipole-coupled heavy neutral lepton (HNL) in MiniBooNE, which directly reflects the example presented in \cref{sec:dipole_examples}.
Here we can see the population of an \cname{InteractionTree} object with a primary interaction ($\nu A \to \mathcal{N} A$) and a single secondary interaction ($\mathcal{N} \to \nu \gamma$).

The fact that every injection distribution is conditional on the previously assigned properties, the detector model, and the set of available interactions, allows for complex injection procedures that can compensate for downstream simulation inefficiencies and acceptance issues.
\siren provides an array of utilities to compute relevant quantities like the ``number of interaction lengths traversed between two points'', that combine information from the physical processes and detector geometry.
These utilities are available directly within injection distributions, which enables users to not only create distributions biased by the detector or interactions, but also to create physically informed distributions that adapt to any detector geometry or choice of interaction model.

Fortuitously, the same features that enable complex injection procedures also afford \siren enormous flexibility in its ability to reweight simulation sets.
Each injection distribution and physics process is required to define both a sampling routine and a way to query the underlying probability density function.
This allows every decision in the injection procedure to be reweighted.
As a result \siren can reweight between different physics models, reweight between different detector geometries, and combine simulation sets generated with disparate injection methods.

Ease of extension by users is one of the main design goals for \siren, and this is most apparent in three categories: physics processes, detector models, and injection distributions.
In general, new physics processes are implemented by subclassing the \cname{CrossSection} or \cname{Decay} classes in \cpp, but we provide a mechanism through \cname{Pybind11}~\cite{pybind11} that allows these classes to be overridden by users in \python.
This allows users to implement new processes quickly without digging into the \cpp internals of \siren.
All detector models are defined via text configuration that describes a list of geometric primitives and density distributions.
This system is flexible enough that most detector configurations can be described to the desired level of accuracy using existing primitives.
However, additional geometric primitives and density distributions will be added as needed.
Finally, the injection distribution can be mixed and matched by users to suit the needs of their simulation scenario.
A variety of distributions of each type are already implemented in \siren, but users can create their own distributions by subclassing either the \cname{PrimaryInjectionDistribution} or the \cname{SecondaryInjectionDistribution} classes.

\siren's codebase is structured as a set of projects that broadly reflect the structure of the event generation procedure described above, but also includes the supporting code that enables these calculations.
These projects are as follows:
\begin{itemize}
    \item \bcname{interactions}: contains the implemented physics processes in addition to the abstract base classes that define their interface.

    \item \bcname{dataclasses}: contains the interaction data structures used throughout \siren.

    \item \bcname{detector}: contains classes for managing the hierarchy of shapes, density distributions, and materials that ultimately define the detector and surrounding material.
        This project also contains utilities for computing interaction lengths and related quantities.

    \item \bcname{distributions}: contains probability distributions that are used for injection and weighting.

    \item \bcname{geometry}: contains definitions of the geometric volumes used to describe the detector geometry components.

    \item \bcname{injection}: contains classes that organize and coordinate the injection procedure and weighting calculation.

    \item \bcname{math}: contains classes that handle various mathematical operations required by \siren, including vector and matrix operations, quaternion operations used to perform rotations, and interpolation methods used to sample from flux and cross section tables.

    \item \bcname{utils}: contains general utilities that may be needed throughout, such as physical constants, error handling, interpolation, and random number generation.
\end{itemize}

Most classes have corresponding \python implementations defined using \cname{pybind11} \python bindings~\cite{pybind11}.
The \python interface to \siren includes an additional helper class, \cname{SIREN\_Controller}, that coordinates setup of the injection procedure.
\cname{SIREN\_Controller} allows the user to specify the experiment under consideration, which determines the input text files for the detector and material models.
It also includes methods to specify the interaction models to be used for event generation and weight calculation, which are not required to be the same.
The user can then use the \cname{GenerateEvents} method of \cname{SIREN\_Controller} to generate a given number of interactions and calculate their physical event weights.
The \cname{SaveEvents} method will then output those events to an Apache Parquet or \cname{HDF5} file using the Awkward Array~\cite{Pivarski:2020qcb} and \cname{h5py}~\cite{andrew_collette_2023_7568214} libraries, a \siren-specific binary format, \cname{siren\_events}.
Additionally, we save the injector object to another \siren-specific binary format, \cname{siren\_injector}, which can then be used for additional event generation or re-weighting purpose.
\Cref{sec:examples} includes examples that further demonstrate the \cname{SIREN\_Controller} interface.

%% file: sections/03_architecture.tex
\section{Architecture}
\label{sec:architecture}

This section touches on the most important components of the \siren architecture in more detail.

\subsection{Dataclasses}


The \bcname{InteractionSignature} class is a simple data structure that holds the types of the initial and final particles in an interaction.
This class is used to uniquely identify types of interactions and is used as a key in the \cname{InteractionCollection} class.
\Cref{table:InteractionSignature} lists the properties of the \cname{InteractionSignature}.
The order of the particles in the signature is important, as it is used to determine the order of the particles in the \cname{InteractionRecord}, and \cname{InteractionRecord} objects with the same set of secondary types but different particle orderings are considered distinct.

\begin{table}
    \begin{tabular}{ll}
        \textbf{Type} & \textbf{Property} \\
        \midrule
        \textbf{ParticleType} & \textmd{primary\_type} \\
        \textbf{ParticleType} & \textmd{target\_type} \\
        \textbf{vector<ParticleType>} & \textmd{secondary\_types} \\
    \end{tabular}
    \caption{Properties of the \texttt{InteractionSignature} class.
    \label{table:InteractionSignature}}
\end{table}

Here \bcname{ParticleType} is simply an \cname{int32\_t} enumeration of different particle types with values that correspond to the PDG particle numbering scheme.


The \bcname{InteractionRecord} class holds all the information needed to describe a single interaction of a primary particle, either via a scattering process off of a target particle or via a decay process.
This includes the types of the initial and final particles, the vertex of the interaction, and the state of these particles before and after the interaction.
\Cref{table:interaction_record} lists the properties of the \cname{InteractionRecord}.
The order of the secondary particles is assumed to be the same as their order in the \cname{InteractionSignature}.
The $\cname{interaction\_vertex}$ is stored in detector coordinates and is necessary for weighting the position probability of the interaction.
The $\cname{initial\_position}$ is also stored in detector coordinates and is necessary for determining the injection bounds for some injection methods.
Finally, the $\cname{interaction\_parameters}$ map is used to store any additional information about the interaction that is not captured by the other properties, but is not used in weighting.

\begin{table}
    \begin{tabular}{ll}
        \textbf{Type} & \textbf{Property} \\
        \midrule
        \textbf{InteractionSignature} & \textmd{signature} \\
        \textbf{ParticleID} & \textmd{primary\_id} \\
        \textbf{array<double, 3>} & \textmd{primary\_initial\_position} \\
        \textbf{double} & \textmd{primary\_mass} \\
        \textbf{array<double, 4>} & \textmd{primary\_momentum} \\
        \textbf{double} & \textmd{primary\_helicity} \\
        \textbf{ParticleID} & \textmd{target\_id} \\
        \textbf{double} & \textmd{target\_mass} \\
        \textbf{double} & \textmd{target\_helicity} \\
        \textbf{array<double, 3>} & \textmd{interaction\_vertex} \\
        \textbf{vector<ParticleID>} & \textmd{secondary\_ids} \\
        \textbf{vector<double>} & \textmd{secondary\_masses} \\
        \textbf{vector<array<double, 4> >} & \textmd{secondary\_momenta} \\
        \textbf{vector<double>} & \textmd{secondary\_helicities} \\
        \textbf{map<string, double>} & \textmd{interaction\_parameters} \\
    \end{tabular}
    \caption{Properties of the \texttt{InteractionRecord} class.}
    \label{table:interaction_record}
\end{table}


The \bcname{InteractionTree} class holds all information about a single event, represented as a tree of \cname{InteractionRecord} objects.
Each node in the tree is a \cname{InteractionTreeDatum} object, which holds a single \cname{InteractionRecord} and a list of its children.
The \cname{InteractionTree} holds a pointer to the primary interaction and a set of pointers to all interactions in the tree.
\Cref{table:interaction_tree} lists the properties of the \cname{InteractionTree} class, and \cref{table:interaction_tree_datum} lists the properties of the \cname{InteractionTreeDatum} class.

\begin{table}
    \begin{tabular}{ll}
        \textbf{Type} & \textbf{Property} \\
        \midrule
        \textbf{InteractionTreeDatum} & \textmd{primary\_interaction} \\
        \textbf{set<InteractionTreeDatum>} & \textmd{tree} \\
    \end{tabular}
    \caption{Properties of the \texttt{InteractionTree} class.}
    \label{table:interaction_tree}
\end{table}

\begin{table}
    \begin{tabular}{ll}
        \textbf{Type} & \textbf{Property} \\
        \midrule
        \textbf{InteractionRecord} & \textmd{record} \\
        \textbf{InteractionTreeDatum} & \textmd{parent} \\
        \textbf{vector<InteractionTreeDatum>} & \textmd{children} \\
    \end{tabular}
    \caption{Properties of the \texttt{InteractionTreeDatum} class.}
    \label{table:interaction_tree_datum}
\end{table}

\subsection{Distributions} \label{sec:distributions}

\begin{figure}
    \centering
\begin{forest}
   for tree={
    align=center,
    s sep'-=7pt,
  },
  forked edges,
    [\cname{WeightableDistribution}
    [\cname{Primary}-\\\cname{InjectionDistribution}
      [Energy]
      [Direction]
      [Helicity]
      [Mass]
      [Vertex]
    ]
    [\cname{Secondary}-\\\cname{InjectionDistribution}
      [Vertex~1D]
    ]
  ]
\end{forest}
    \caption{
    Class hierarchy of the \textit{distributions} project.
    The \texttt{WeightableDistribution} class is an abstract base class used to represent any distribution that can be used for weighting.
    The \texttt{PrimaryInjectionDistribution} and \texttt{SecondaryInjectionDistribution} classes are used to define the distributions from which primary and secondary particle properties are sampled, respectively.
    These three classes are fundamental to the injection and weighting architecture of \siren.
    Further subclasses are used to define common types of distributions but are not fundamental to the architecture.
    }\label{fig:distributions_hierarchy}
\end{figure}
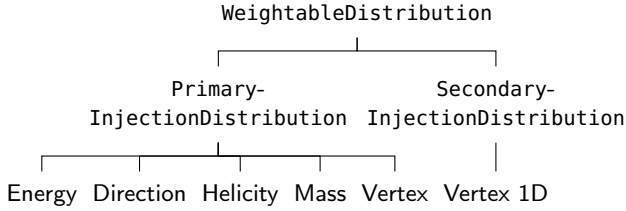

The probability distributions from which interaction properties are sampled and the physical distributions that interactions are weighted to are defined within the \cname{distributions} project.
These distributions follow the class hierarchy outlined in \cref{fig:distributions_hierarchy}.
This hierarchy starts with the abstract base class \bcname{WeightableDistribution} which represents any distribution that can be used for weighting.
\cname{WeightableDistribution} requires implementing a \cname{GenerationProbability} method that returns the probability density for an interaction to have been sampled from this distribution.
This method takes an \cname{InteractionRecord} as its first argument, which describes the properties of the interaction, but also takes a \cname{DetectorModel} and \cname{InteractionCollection} object which can be used as contextual information to modify the probability distribution.

The two subclasses \bcname{PrimaryInjectionDistribution} and \bcname{SecondaryInjectionDistribution} are used to define the distributions from which primary and secondary particle properties are sampled, respectively.
The main difference between these two classes is that all properties of a primary particle must eventually be sampled by a distribution, while only the interaction vertex of a secondary particle must be sampled, as the kinematic properties of the secondary particle are fixed by the previous interaction.
Both classes require the implementation of a \cname{Sample} method that populates the relevant properties of an \cname{InteractionRecord} object, and a \cname{GenerationProbability} method that returns the probability density of the distribution.
These \cname{Sample} methods respectively take a \cname{PrimaryDistributionRecord} or \cname{SecondaryDistributionRecord} object as input, which provide a high-level interface to access and assign properties of the primary and secondary particles, respectively.
In the case that an unset property of the interaction is accessed through either of these interface classes, the property is either computed from available information, or an informative error is thrown.

\subsection{Interactions} \label{sec:interactions}

\siren supports both $2 \to n$ and $1 \to n$ processes which are described by the \cname{CrossSection} and \cname{Decay} classes, respectively.
The \cname{interactions} project contains these abstract base classes and provides implementations for the processes described in \cref{sec:provided_xsec,sec:provided_dec}.

The \bcname{CrossSection} class requires the implementation of several methods; of these \cname{TotalCrossSection}, \cname{DifferentialCrossSection}, and \cname{FinalStateProbability} are used in the weighting of interactions as well injection techniques that depend on interaction probability.
Beyond these methods, subclasses of \cname{CrossSection} must also implement a \cname{SampleFinalState} method that samples the secondary particle properties; this method takes a \cname{CrossSectionDistributionRecord}.
In a similar manner to the \cname{Primary-} and \cname{Secondary-DistributionRecord}, the \cname{CrossSectionDistributionRecord} provides an interface for querying properties of the interaction and setting properties of the secondary particles.

The \bcname{Decay} class has an almost identical interface to \cname{CrossSection}, except that the cross section-specific methods are replaced by \cname{TotalDecayWidth} and \cname{DifferentialDecayWidth}, which are also used in weighting and interaction probability-based injection.

\subsection{Detector Interface} \label{sec:detector_interface}

The \bcname{DetectorModel} class provides an interface for computing things like the material density, column depth, number of interaction lengths between points, and other quantities that depend on the geometry and material composition of the detector.
A \cname{DetectorModel} instance contains a full description of the geometry, density distribution, and material composition of the detector to facilitate these calculations.

The geometry of the detector is described by a hierarchy of \bcname{DetectorSector} objects.
Each \cname{DetectorSector} object has the properties: name, material identifier, hierarchy level, geometric shape, and density distribution.
The hierarchy level is unique to each sector and determines precedence in the case of overlapping sectors.
A simple tracking algorithm allows a callback function to be called within a loop over the relevant segments of individual sectors.
This tracking algorithm and callback function system is used to compute integrals and other quantities across a path through the detector.

\begin{table}
    \begin{tabular}{ll}
        \textbf{Type} & \textbf{Property} \\
        \midrule
        \textbf{string} & \textmd{name} \\
        \textbf{int} & \textmd{material\_id} \\
        \textbf{int} & \textmd{level} \\
        \textbf{shared\_ptr<Geometry>} & \textmd{geo} \\
        \textbf{shared\_ptr<DensityDistribution>} & \textmd{density} \\
    \end{tabular}
    \caption{Properties of the \texttt{DetectorSector} struct}
    \label{table:detector_sector}
\end{table}

The geometric shape of each \cname{DetectorSector} is described by an instance of the \bcname{Geometry} class, an abstract base class that requires the implementation of a \cname{ComputeIntersections} method to compute the intersections of a ray with the shape's boundary, assuming an a-priori defined orientation of the shape.
At the time of writing the implemented shapes include a \bcname{Box}, \bcname{Cylinder}, \bcname{Sphere}, and \bcname{ExtrudedPolygon}; a list that will be expanded upon in the future to accommodate the proper description of new experiments.
Beyond the parameters of any individual geometric shape, each \cname{Geometry} object has both a name and \cname{Placement}, which describes the position of and orientation of the geometric shape within the global geometry coordinate system.
The \cname{Placement} object is used to perform transformations between the global geometric coordinate system and the a-priori-defined local coordinate system of the shape when intersections and other quantities are computed.

\begin{table}
    \begin{tabular}{ll}
        \textbf{Type} & \textbf{Property} \\
        \midrule
        \textbf{string} & \textmd{name} \\
        \textbf{Placement} & \textmd{placement} \\
    \end{tabular}
    \caption{Properties of the \texttt{Geometry} class}
    \label{table:detector_sector}
\end{table}

\begin{table}
    \begin{tabular}{ll}
        \textbf{Type} & \textbf{Property} \\
        \midrule
        \textbf{Vector3D} & \textmd{position} \\
        \textbf{Quaternion} & \textmd{rotation} \\
    \end{tabular}
    \caption{Properties of the \texttt{Placement} class}
    \label{table:detector_sector}
\end{table}

Each \cname{DetectorSector} has an integer material identifier and \cname{DensityDistribution} to describe its atomic composition and density, respectively.
The \bcname{DensityDistribution} abstract base class requires the implementation of several methods related to density calculations:

\begin{itemize}
    \item \cname{Evaluate}: value of the density at a single point
    \item \cname{Derivative}: one-dimensional directional-derivative of the density at a single point
    \item \cname{AntiDerivative}: indefinite one-dimensional directional-integral of the density at a single point
    \item \cname{Integral}: definite one-dimensional directional-integral of the density between two points
    \item \cname{InverseIntegral}: distance from a single point along a path for which a one-dimensional directional-integral between the points is equal to a particular value
\end{itemize}

These methods are assumed to operate within the global geometry coordinate system and to be self-consistent within numerical tolerance.

The \bcname{DensityDistribution1D} template class inherits from \cname{DensityDistribution} and its specializations provide a variety of common distributions that be can described as one-dimensional functions.
Specializations of \cname{DensityDistribution1D} depend on two types that are themselves sub-classes of the  \cname{Axis1D} and \cname{Distribution1D} abstract base classes.
The \bcname{Axis1D} class has a position and direction that are used to define the orientation of the one-dimensional axis, and sub-classes are required to implement \cname{GetX} and \cname{GetdX} methods that return the 1D position on the axis that is occupied by a 3D point in the global geometric coordinate system, and the Jacobian factor between the distance along a ray in 3D space and a distance along the 1D axis, respectively.
At the time of writing two \cname{Axis1D} subclasses are available, the \bcname{CartesianAxis1D} which defines an axis along a single direction in 3D space, and the \bcname{RadialAxis1D} which defines a 1D axis along the radial direction from some central point.
The \bcname{Distribution1D} abstract base class requires the implementation of \cname{Evaluate}, \cname{Derivative}, and \cname{AntiDerivative} methods that correspond to the 1D versions of such functions defined for \cname{DensityDistribution}.
At the time of writing three \cname{Distribution1D} specializations are available, the \cname{ConstantDensityDistribution}, \cname{ExponentialDensityDistribution}, and \cname{PolynomialDensityDistribution}.
\cname{DensityDistirbution1D} provides generic default implementations of the methods required by \cname{DensityDistribution}, but specializations for certain \cname{Axis1D} and \cname{Distribution1D} combinations are implemented for improved performance.

The \bcname{MaterialModel} class manages material definitions and provides methods for querying aggregate material properties.
The injection methods and weighting calculations of \siren are primarily concerned with atomic targets, protons, neutrons, and electrons, so the chemical composition of materials is neglected and only the isotopic composition is tracked.
Each material is defined by a series of PDG identification numbers for each isotope within the material and a corresponding fraction by weight for each isotope.
At the time of writing a table of atomic masses for each isotope from~\cite{Wang:2021xhn} is hard coded within the \cname{MaterialModel}, and the masses of both hyper-nuclei and isotopes missing from the table are approximated by subtracting an empirical estimate of the nuclear binding energy~\cite{Samanta:2005kd} from the sum of constituent masses.

The \cname{DetectorModel} uses two coordinate systems, not including the local coordinate systems of the \cname{Geometry} objects: a global geometry coordinate system we refer to as \cname{GeometryCoordinates}, and a detector-specific coordinate system that we refer to as \cname{DetectorCoordinates}.
\cname{GeometryCoordinates} serve as a global coordinate system that can be used to define a detector hall or other large features common to different experiments.
This coordinate system is used internally for all calculations within the \cname{DetectorModel}; most methods of \cname{DetectorModel} that use \cname{DetectorCoordinates} are private.
In contrast, \cname{DetectorCoordinates} serve as a more convenient detector-specific coordinate system that is used within the injection distributions and intended for compatibility with analyses.
This coordinate system is also used by the \cname{InteractionRecord} objects and the majority of the public \cname{DetectorModel} methods.
Transformations between the coordinate systems are defined by a position and rotation stored within \cname{DetectorModel}.
To reduce the possibility of errors confusing the two coordinate systems, four "strong-types" are defined using the \cname{fluent::NamedType} package~\cite{FluentNamedtype}, namely: \cname{GeometryPosition}, \cname{GeometryDirection}, \cname{DetectorPosition}, and \cname{DetectorDirection}.
These strong-types are essentially structs that encapsulate a \cname{Vector3D} object and have convenience functions to expose operators and methods of the underlying type.
Since these types are not implicitly convertible between one another or the underlying \cname{Vector3D} type, the result of a calculation in one coordinate system cannot easily be passed to a function expecting the other coordinate system.
To simplify the general user experience both types are exposed to \python, but implicit conversions to and from \cname{Vector3D} are enabled only for the \cname{DetectorCoordinates} types.

Many of the vertex injection distributions first choose a fixed path through the detector, and then query properties of the detector and interaction with respect to this path before choosing the final interaction vertex location.
The \bcname{Path} class is provided to facilitate these calculations and manipulations of a path through the detector.
One can, for example, define a \cname{Path} with a single starting point, direction, and length, then extend the \cname{Path} by a number of interaction lengths, before finally bounding the \cname{Path} to the finite volume.
Each of these three steps can be performed with a single function call: \cname{Path::Path}, \cname{Path::ExtendFromStartByInteractionDepth}, and \cname{Path::ClipToOuterBounds}.
Beyond providing a convenient interface for manipulating paths through the detector, the \cname{Path} class also provides some performance improvements by caching the computed list of \cname{Intersections} with the detector geometry.

%% file: sections/04_features.tex
\section{Features}
\label{sec:features}

This section describes the features that make \siren unique within the community of Monte Carlo tools for neutrino physics.
In general, \siren connects new physics models and the detailed geometric features of neutrino detectors, enabling fast and accurate simulation of rare neutrino interactions.
This is reflected in the features described below, which include (1) a generalized interaction interface, (2) a detailed geometric interface, (3) support for multiple injection methodologies, and (4) exceptional computational performance.

\subsection{Generalized Interaction Interface}

\siren is designed to support a variety of interaction models.
These are grouped into two categories, cross sections and decays, which represent $2 \to n$ and $1 \to n$ interactions, respectively; where $n$ is the number of final state particles.
We have included several cross section and decay models with \siren, as well as an interface with the \darknews software package to support the new physics models provided there.
Additional interactions can be included as \python subclasses of the \cname{CrossSection} or \cname{Decay} \cpp classes.
As discussed in \cref{sec:interactions}, one needs to implement and override a set of methods enabling \siren to evaluate total and differential cross sections or decay widths and determine the kinematics of final state particles in the interaction.
The suite of interaction models provided directly within \siren is intended to grow over time.
Below we describe the interaction model implementations provided with \siren at the time of writing.

\subsubsection{\darknews Compatibility}
\label{sec:dark_news}

In order to significantly extend the set of models supported by \siren, we have implemented a custom interface with the \darknews software package~\cite{Abdullahi:2022cdw}.
\darknews calculates cross sections and decay widths for interactions between the three active neutrinos $\{\nu_e, \nu_\mu, \nu_\tau\}$ and three additional heavy neutral leptons (HNLs) $\{N_4,N_5,N_6\}$, mediated by the SM $Z$ boson, the SM photon $(\gamma)$, a new dark photon $(Z')$, and/or a new dark scalar $(h')$.
The phenomenological signatures of these models typically involve the production of the HNLs via upscattering off of nuclear targets followed by the decay of the HNLs to visible particles.
\darknews uses the \texttt{vegas} algorithm~\cite{Lepage:1977sw} to calculate event rates in a detector volume via a multi-dimensional integration of the differential cross section and decay width.
The \siren built-in \darknews interface enables the user to simulate these models in arbitrary detector geometries.
This is important, as the spatially extended nature of the production-and-decay signature means that event rates and kinematic distributions can depend strongly on the exact geometry of the detector and surrounding environment.

The interface consists of two \cpp abstract classes, \cname{DarkNewsCrossSection} and \cname{DarkNewsDecay}, which have corresponding \python derived classes within the \texttt{python/SIRENDarkNews.py} file.
This file also includes the \cname{PyDarkNewsInteractionCollection} class, which collects the different available cross sections and decays for a given \darknews model.
Instances of this class are defined by passing a dictionary of physics parameters that employs the same format as the official \darknews python interface.

\siren considers each interaction vertex of an event separately; thus, we cannot integrate over the differential cross section and decay width simultaneously to determine the overall event rate as in \darknews.
Instead, we treat the cross sections and decays of a given \darknews model separately.
The \darknews cross section interface builds up tables to store the total and differential cross section as a function of $E_\nu$ and $\{E_\nu, z\}$, respectively, where
\begin{equation}
z \equiv \frac{Q^2 - Q^2_{\rm min}(E_\nu)}{Q^2_{\rm max}(E_\nu) - Q^2_{\rm min}(E_\nu)},
\end{equation}
and $Q^2_{\rm min/max}(E_\nu)$ are the energy-dependent minimum and maximum four momentum transfer of the process.
The lowest energy node of each table corresponds to the energy threshold of the process at hand.
The table entries are then pre-determined according to a user-defined interpolation tolerance $\alpha$ with a default value of 4\%.
The entries of the total cross section table are given by the set
\begin{equation}
\{(1+\alpha)^i (1+\beta) E_\nu^{\rm thresh}\}_i,
\end{equation}
where $E_\nu^{\rm thresh}$ is the energy threshold of the interaction, $\beta$ is a user-defined numerical tolerance with a default value of $10^{-6}$, and $i \in \mathbb{Z}$.
For the differential cross section, table entries are given by
\begin{equation}
\{(1+\alpha)^i (1+\beta) E_\nu^{\rm thresh},~(1 +\alpha)^{j} \beta \}_{i,j},
\end{equation}
where $i,j\in\mathbb{Z}$ and $j$ is bounded from above such that the $z$ entries of the table do not exceed unity.
When the simulation requests a total or differential cross section at a point outside the existing table bounds, \siren will query \darknews for table entries from the bottom up until the requested point is within table bounds.
The requested cross section value is then given by interpolating between points in the table.
We use the \cname{PchipInterpolator} (\cname{LinearNDInterpolator}) method of the \cname{scipy.interpolate} library for the total (differential) cross section table~\cite{2020SciPy-NMeth}.
This method ensures reproducibility and reasonable computational efficiency once the tables have been populated.
The user can choose to either pre-compute cross section tables out to a specified maximum energy, or to iteratively-generate cross section tables as different energy points are requested.
Furthermore, \siren supports serialization of the tables, which means that simulations can be re-weighted to updated versions of these models, from \darknews or otherwise, that result in changes to computed cross section values.

The \darknews decay interface does not require interpolation of tables.
Instead, we use the \darknews internal methods to compute the differential and total decay width in the rest frame of the decaying particle.
The exact variables used to parameterize the phase space depend on the decay model and are always sufficient to fix the kinematics of the outgoing decay products~\cite{Abdullahi:2022cdw}.
When simulating a \darknews decay in \siren, we use the \texttt{vegas}-based \darknews internal methods to sample the phase space of the rest frame differential decay width.
Specifically, we store a set of phase space points and their corresponding probabilities, such that we can sample from the cumulative distribution function of this set to determine the kinematics of final state particles.
\siren supports serialization of the \darknews \texttt{vegas} integrator and phase space samples/probabilities, which also ensures reproducibility and enables re-weighting to future \darknews versions.

\subsubsection{Currently available cross sections} \label{sec:provided_xsec}

\siren provides several out-of-the-box cross section models ready for use, including
\begin{itemize}
    \item \textbf{Neutrino DIS:} $\nu_\alpha N \to \ell_\alpha X$ (CC) and $\nu_\alpha N \to \nu_\alpha X$ (NC).
    This model describes neutrino interactions with the constituent quarks inside a nucleon $N$ to produce an outgoing lepton $\nu_\alpha$/$\ell_\alpha$ and hadronic system $X$.
    Total and differential cross sections are provided as splines computed using the \texttt{photospline} software package~\cite{Whitehorn:2013nh}.
    These splines are based on the CSMS neutrino DIS cross section calculation~\cite{Cooper-Sarkar:2011jtt}, and the Metropolis-Hastings method~\cite{Metropolis:1953am,Hastings:1970aa} is used to sample the double differential cross section in Bjorken $x$ and $y$.

    \item \textbf{Neutrino dipole-portal upscattering:} $\nu_\alpha A \to N A$ (coherent) and $\nu_\alpha p \to N p$ (inelastic) upscattering via a transition magnetic moment.
    This model describes neutrino interactions with a nucleus $A$ or proton $p$ to produce an HNL $N$.
    Total and differential cross sections are calculated by interpolating pre-computed tables.
    These tables have historically been generated using \darknews, with the single differential cross section computed as a function of Bjorken $y$ or $z \equiv (y-y_{\rm min}) / (y_{\rm max} - y_{\rm min})$.
    The final state kinematics of the outgoing HNL and nuclear system are sampled using the Metropolis-Hasting method.
    This class has since been superseded by the dedicated \darknews cross section interface of \siren described in \cref{sec:dark_news}; therefore, cross sections tables for this class are not provided directly in \siren.

    \item \textbf{Neutrino-electron elastic scattering:} $\nu_\alpha e^- \to \nu_\alpha e^-$.
    This model describes elastic collisions between neutrinos and electrons.
    We use the tree-level differential cross section as a function of Bjorken $y$ in Appendix~A of Ref.~\cite{Valencia:2019mkf}, with electroweak couplings from Ref.~\cite{Erler:2013xha}.
    The total cross section is computed using Romberg integration.
    The final state kinematics of the outgoing neutrino and electron are sampled using the Metropolis-Hasting method.

    \item \textbf{\darknews cross sections:} $\nu_\alpha A \to N A$ (coherent) and $\nu_\alpha n \to N n$ (inelastic) via a mediator in $\{Z,\gamma,Z',h'\}$.
    These models describe the upscattering of active neutrinos off of a nucleus $A$ or nucleon $n$ into HNL states through the mediators described in \cref{sec:dark_news}.
    Total differential cross sections are calculated by interpolating tables generated using \darknews.
    The final state kinematics of the outgoing HNL and nuclear system are sampled using the Metropolis-Hasting method alongside the single differential cross section as a function of $z$ as defined in \cref{sec:dark_news}.
    
\end{itemize}

\subsubsection{Currently available decays} \label{sec:provided_dec}

\siren also provides a few out-of-the-box decay models ready for use, including

\begin{itemize}

    \item \textbf{HNL dipole-portal decays:} $N \to \nu \gamma$ via a transition magnetic moment.
    This model describes the single photon decay of an HNL via a transition magnetic moment operator.
    The total and differential decay widths are computed as described in Ref.~\cite{Kamp:2022bpt}, which can vary depending on the Dirac or Majorana nature of the HNL.
    This class has since been superseded by the dedicated \darknews decay interface of \siren described in \cref{sec:dark_news}.

    \item \textbf{\darknews decays:} $N \to \nu X$, where $X \in \{e^+ e^-, \mu^+ \mu^-, \gamma\}$ via a mediator in $\{Z,\gamma,Z',h'\}$.
    This model describes the visible decays of HNLs through the mediators described in \cref{sec:dark_news}.
    Total and differential cross sections are computed using \darknews methods.
    The final state kinematics of the outgoing particles are determined by sampling the phase space of the differential decay width, which depends on the decay model under consideration as described in \cref{sec:dark_news}.

\end{itemize}

\subsection{Detailed Geometry Configurations}

The power of \siren is contained in its ability to simulate arbitrarily complex detector geometries.
This is accomplished using the detailed detector interface described in \cref{sec:detector_interface}.
This text-based input allows the user to easily define a series of detector sectors that describe the full detector geometry.
This gives the user control over the level of detail with which to describe the detector geometry, which is always a trade-off between accuracy and computational complexity.
The detector interface of \siren is important for the robust phenomenological evaluation of new physics models in a given experiment.
Like the interaction models, the suite of detector models provided directly within \siren is intended to grow over time.
This section touches on the detector models already provided within \siren at the time of writing.
The level of detail for these models reflects that which was necessary for existing studies.
It is important to note that the detector models are version-controlled, and can therefore be consistently updated with increased detail.

\subsubsection{Currently available geometries} \label{sec:provided_detectors}

\siren provides several out-of-the-box geometry models ready for use, including

\begin{itemize}

    \item \textbf{IceCube:} As the IceCube detector is sensitive to neutrino interactions occurring in the ice, Earth, and atmosphere, we use the Preliminary Reference Earth Model (PREM)~\cite{Dziewonski:1981xy} to describe the IceCube detector environment.
    This model is extended with three additional uniform density layers representing clear ice, firn (an intermediate stage between snow and ice), and the atmosphere~\cite{IceCube:2020tcq}.
    The IceCube detector is taken to be a cylinder of ice with a radius of 546.2~${\rm m}$ and height of 1000~${\rm m}$ (corresponding to a volume of 1~${\rm km}^{3}$) situated at a depth of 1450-2450~${\rm m}$ below the surface of the ice~\cite{IceCube:2016zyt}.
    
    \item \textbf{DUNE:} The DUNE detector was first considered in \siren within the context of atmospheric neutrinos~\cite{Schneider:2021wzs}.
    Therefore, we also implement DUNE within the PREM description of the Earth, with an additional constant-density atmosphere layer.
    We consider a single DUNE far detector module, which consists of a $14\;{\rm m} \times 58.2\;{\rm m} \times 12\;{\rm m}$ rectangular prism of liquid argon situated 1480~${\rm m}$ below the Earth's surface~\cite{DUNE:2020txw} to model interactions within the liquid argon, but neglect a description of the material in the surrounding hall.
    
    \item \textbf{ATLAS:} The ATLAS detector was first considered in \siren within the context of higher energy supernova neutrinos~\cite{Wen:2023ijf}.
    We implement ATLAS within the same PREM and atmosphere description of the DUNE detector model.
    The hadronic calorimeter and muon spectrometer components of the ATLAS detector are represented as concentric cylinders embedded 90~m below the surface of the Earth.
    We model the hadronic calorimeter as an iron cylinder approximately 12.3~${\rm m}$ long with inner and outer radii of approximately 2.3~${\rm m}$ and 3.8~${\rm m}$, respectively \cite{ATLAS:1996aa}.
    Because the study concerned~\cite{Wen:2023ijf} did not consider interactions within the muon system it is approximated as a cylinder of length 40~${\rm m}$ and radius 11~${\rm m}$, which are the general outer dimensions of the ATLAS detector itself \cite{ATLAS:2008xda}.
    
    \item \textbf{Hyper-K:} We also model Hyper-K using the PREM and atmosphere description of the Earth described above.
    We consider the Hyper-K detector to be a cylinder of water with a radius of 34~${\rm m}$ and height of 60~${\rm m}$~\cite{Hyper-Kamiokande:2018ofw}.
    The detector is situated the detector 650~m below the surface of the Earth.

    \item \textbf{MiniBooNE:} We model MiniBooNE as a 6.1~m radius sphere of mineral oil~\cite{MiniBooNE:2008paa}, with a fiducial radius of 5~m~\cite{MiniBooNE:2020pnu}.
    MiniBooNE is surrounded by an additional 3~m of air representing the detector hall as well as a 600~m-long rectangular prism of bedrock representing the path of the Booster Neutrino Beamline (BNB) (though the maximum generation distance of neutrino interactions can be fixed to match the physical BNB target distance of 541~m).
    We define bedrock as SiO$_2$ with a uniform density of 2.9~\si{\g \per \cm^3}, though the user can easily change the composition and density.
    
    \item \textbf{MINER$\mathbf{\nu}$A:} The \minerva detector is considerably more complicated than MiniBooNE, and an accurate description was necessary to model dipole-coupled HNL production via upscattering in Ref.~\cite{Kamp:2022bpt}.
    We follow Ref.~\cite{MINERvA:2013zvz} to construct our \minerva model, including the scintillator planes of the inner tracker, the lead, steel, and carbon nuclear target region, the surrounding lead and scintillator-based electromagnetic calorimeter, and the upstream steel veto shield.
    All components of the \minerva detector are implemented using extruded polygons, allowing an accurate description of the overall hexagonal prism structure and the irregular extruded polygon shapes comprising the nuclear targets.
    The fiducial volume of \minerva is a 5.99 metric ton subset of the active tracker region defined by a hexagonal prism with an apothem of 81.125\;cm~\cite{Valencia:2019mkf}.
    The detector is situated inside a 10~m sphere of air representing the detector hall as well as a 300~m-long rectangular prism of bedrock representing the path of the Neutrino Main Injector (NuMI) beamline (though again, the maximum generation distance can be tuned to the actual NuMI bedrock transit distance of 240~m).
    
    \item \textbf{CCM:} The CCM detector consists of a 0.96~m-radius and 1.232~m-height cylinder of liquid argon operating at the Lujan beam dump facility of the Los Alamos Neutron Science Center~\cite{CCM:2021leg}.
    While the detector itself is relatively simple, the detector environment is relatively complicated due to the extensive shielding between CCM and the Lujan tungsten target.
    We model the most relevant parts of this environment, which is necessary for models in which the production of new particles happens in this shielding (including the dipole-portal example discussed in \cref{sec:dipole_examples}).
    This includes the lead, steel, and beryllium target-moderator-reflector-shield (TMRS)~\cite{ZAVORKA2018189}, the steel shielding surrounding the TMRS, additional walls of concrete, steel, and borated polyethylene shielding constructed between the target and CCM, the CCM cryostat, and liquid argon active volume~\cite{EdThesis}.
    The entire detector environment is placed on top of an 8~m deep rectangular prism of concrete and within a larger rectangular prism of air, representing the Lujan floor and detector hall, respectively.
    
\end{itemize}

\subsection{Tabulated Flux Tables} \label{sec:flux_tables}

\siren is designed to support the injection of neutrino events within a wide variety of experimental configurations.
This requires the ability to sample from and reweight to neutrino flux models that do not necessarily have clean analytic expressions.
We have included a \cname{TabulatedFluxDistribution} class to address this.
Here, the user provides tabulated data describing the neutrino flux as a function of energy, which \siren then uses to construct the neutrino energy probability density function and sample neutrinos from the inverse cumulative distribution function.
The user can also indicate whether the provided flux table has a physical normalization, that is, whether the neutrino flux entries of the table contain information about the total neutrino flux.
If this is the case, the table entries should reflect the distribution,
\begin{equation}
\frac{d\phi}{dE dA dt}~\bigg[\frac{\nu}{{\rm GeV} {\rm m}^2 {\rm T}}\bigg],
\end{equation}
where T is a unit describing the livetime of the experiment, e.g. protons-on-target (POT) or years.
When these flux units are employed, \siren will calculate event weights in units of ${\rm T}^{-1}$.

In order for users to get started, we have provided a few flux tables packaged with \siren that can be used immediately.
At the time of writing, these include

\begin{itemize}

    \item \textbf{BNB}: We provide tables of the Booster Neutrino Beam (BNB) $\nu_\mu$, $\nu_e$, $\bar{\nu}_\mu$, $\bar{\nu}_e$ flux in both forward horn current (FHC) and reverse horn current (RHC) mode.
    The flux is given with respect to a single proton-on-target (POT).
    The BNB flux calculation comes from Ref.~\cite{MiniBooNE:2008hfu}.

    \item \textbf{NuMI}: We provide tables of the Neutrino Main Injector (NuMI) $\nu_\mu$, $\nu_e$, $\bar{\nu}_\mu$, $\bar{\nu}_e$ flux in FHC and RHC mode for the low energy (LE) and medium energy (ME) configurations of NuMI.
    The LE flux comes from the \minerva data release associated with Ref.~\cite{MINERvA:2016iqn}, and the ME flux comes from Ref.~\cite{AliagaSoplin:2016shs}.
    As in the BNB case, the flux is given with respect to a single POT.
    This is the same treatment of the NuMI flux employed by \darknews~\cite{Abdullahi:2022cdw}.

    \item \textbf{HE SN}: We provide an example high energy supernova (SN) neutrino flux computed in Ref.~\cite{Murase:2017pfe}, corresponding to the ATLAS example presented in Ref.~\cref{sec:DIS_examples}.
    Specifically, \siren provides a table of the $\nu_\mu$ flux from a Type \RNum{2}n SN explosion at a distance of 10 \si{\kilo\parsec} over a period of 100 days.
    
\end{itemize}

\subsection{Injection Methodologies} \label{sec:injection_methodologies}

\siren separates the injection of physics interactions from the calculation of their physical weights.
Thus, the user can typically decide from among multiple injection schemes for a single simulation task.
Different physics scenarios may lend themselves to one particular injection scheme over another.
For example, if one is interested in determining the sensitivity of IceCube to $\nu_e$ CC interactions within the detector, it is more efficient to inject interaction vertices uniformly throughout the detector.
However, if one is instead interested in the rate of through-going muons from $\nu_\mu$ interactions passing through the DUNE volume, one must now consider interactions outside of the detector.
In this case, it is more efficient to use a ranged injection scheme in which the maximum distance of interaction vertices from the detector is determined on an event-by-event basis from the maximum range of the outgoing muon.
Furthermore, suppose one is interested in a model in which HNLs are produced outside the MiniBooNE detector but decay within the fiducial volume.
In that case, it is more efficient to use a different ranged injection scheme in which the range is determined by the outgoing HNL lifetime.
For this reason, \siren includes support for a variety of injection schemes through the \texttt{distributions/primary/vertex/} project.
The classes implemented here can be thought of as a special set of distributions determining the location of the initial interaction.

The different injection schemes of \siren can currently be grouped into two categories: \textbf{volume injection}, in which interaction locations are sampled uniformly within a specified three-dimensional volume, or \textbf{ranged injection}, in which interaction locations are sampled according to some specified probability density along the line of sight of the neutrino, within a specified range.
The former is useful when the detector is only sensitive to primary interactions that occur within the fiducial volume, and the latter is useful when the detector is sensitive to particles produced in interactions that occur outside the fiducial volume.
These categories are inspired by \prevleptoninjector but have been extended to accommodate physics scenarios beyond neutrino DIS at IceCube, and can be readily extended by users to accommodate other scenarios.

The different primary interaction position distributions provided by \siren include

\begin{itemize}

    \item \bcname{CylinderVolumePositionDistribution}: Neutrino interactions are injected uniformly within a cylinder of specified inner/outer radius, height, and position within the detector model.
    The \cname{OrientedCylinderVolumePositionDistribution} class operates similarly but supports rotations of the generation volume.

    \item \bcname{RangePositionDistribution}: The location of the neutrino interaction is determined by first sampling a point of closest approach uniformly from a disk of fixed radius centered at the detector and oriented origin perpendicular to the neutrino direction.
    The path of the neutrino is determined by extending a line segment from the point of closest approach by a fixed ``endcap length'' in both directions perpendicular to the plane of the disk, i.e. along the neutrino direction.
    This line segment is extended further upstream by a specified range, which can depend on the properties of the initial neutrino.
    The neutrino location is then determined by sampling a position along the line segment according to the traversed interaction depth, which depends on the specific interactions available to the neutrino as well as the traversed detector components.
    This setup follows closely from \prevleptoninjector \cite{IceCube:2016zyt}.
    
    \item \bcname{ColumnDepthPositionDistribution}: Neutrino interactions are sampled similarly to the \cname{RangePositionDistribution} case; however, as this method is intended for neutrino DIS, the range corresponds to the maximum survival distance of the outgoing $\mu$ or $\tau$.
    Furthermore, the neutrino interaction location along the path is sampled according to the traversed column depth rather than interaction depth, which does not require evaluations of the total cross section and thus has improved computational efficiency.

    \item \bcname{DecayRangePositionDistribution}: Neutrino interactions are sampled similarly to the \cname{RangePositionDistribution} case; however, the range here corresponds to the decay length of the final state particle of interest.
    Furthermore, the neutrino interaction location is also sampled according to the decay length.
    This is intended to efficiently sample events that result in an observable decay within the fiducial volume.
    It is worth noting that this is the only ranged injection scenario in which interaction locations are not sampled physically along the neutrino line of sight, though this is correctly accounted for in the weight calculation.

    \item \bcname{PointSourcePositionDistribution}: This is a somewhat unique injection methodology in which neutrino interactions are sampled along the line of sight from a specific point in the detector model.
    The interaction location along the path is sampled according to the interaction depth of the neutrino.
    This scheme is intended for use in experiments for which neutrinos come from a single point source, but the detector is close enough that the neutrino direction no longer follows a plane wave approximation (e.g., the CCM example in \cref{sec:dipole_examples}).
    This position distribution comes with a word of caution: weights calculated while using this position distribution will not include a factor of $m^{-2}$, which means the total flux normalization from the source must be provided in units of $[\nu s^{-1}]$ for consistent event rate calculations.
    
\end{itemize}

\subsection{Computational Performance}

\siren is intended to be lightweight and enough to be useful for phenomenological studies.
Computational efficiency is thus an essential feature of the software package.
It is generally true that the more detailed the detector model and/or interaction model, the more computationally expensive the simulation.
The detector model methods described in \cref{sec:detector_interface} will be used in any \siren simulation job and are thus designed with efficiency in mind.
In contrast, the interaction models are more specific to each use case and can be implemented by the user.
That being said, the methods of the \siren-provided interaction models described in
\cref{sec:provided_xsec,sec:provided_dec} are also designed with efficiency in mind.
Even in the most complicated cases, \siren can still generate $\mathcal{O}(100-1000)$ events per second.
This is shown in \cref{tab:computation_efficiency}, which reports the time required to generate events and calculate weights for the examples presented in \cref{sec:examples}.
The distributions used to generated \cref{tab:computation_efficiency} are provided in \cref{app:comp_efficiency}.

From \cref{tab:computation_efficiency}, one can see that the DIS examples are more efficient than the HNL examples, which is consistent with the simpler interaction model.
The IceCube and DUNE DIS examples, which use ranged injection, have a slightly longer event generation time compared to the ATLAS example, which uses volume injection.
This is also true, albeit to a lesser extent, for the weight calculation time.
Event generation and weight calculation are more expensive in IceCube than in DUNE, likely due to the computational difference between calculating intersection with a cylinder and a rectangular prism.
Overall, weight calculation is slightly more expensive than event generation for all of the DIS examples, likely because the computational efficiency here is driven by sampling the interaction location.

For the HNL examples, event generation times in \minerva are longer and more variable than in the MiniBooNE and CCM cases.
This is a consequence of the more complicated detector geometry and higher typical neutrino energies, the latter of which leads to larger cross section tables.
The MiniBooNE example is reasonably computationally efficient due to the simple detector model and lower typical neutrino energies. 
The CCM case is also reasonably computationally efficient despite the complicated detector environment.
This is because neutrinos are mono-energetic in this example, leading to fast evaluations of total and differential cross sections.
Weight calculation is less expensive than event generation in the HNL examples; this is driven by the Metropolis-Hastings sampling of the differential cross sections in the \cname{DarkNews} interface during the event generation.

\begin{figure*}[h]
    \centering
    \includegraphics[width=0.9\textwidth]{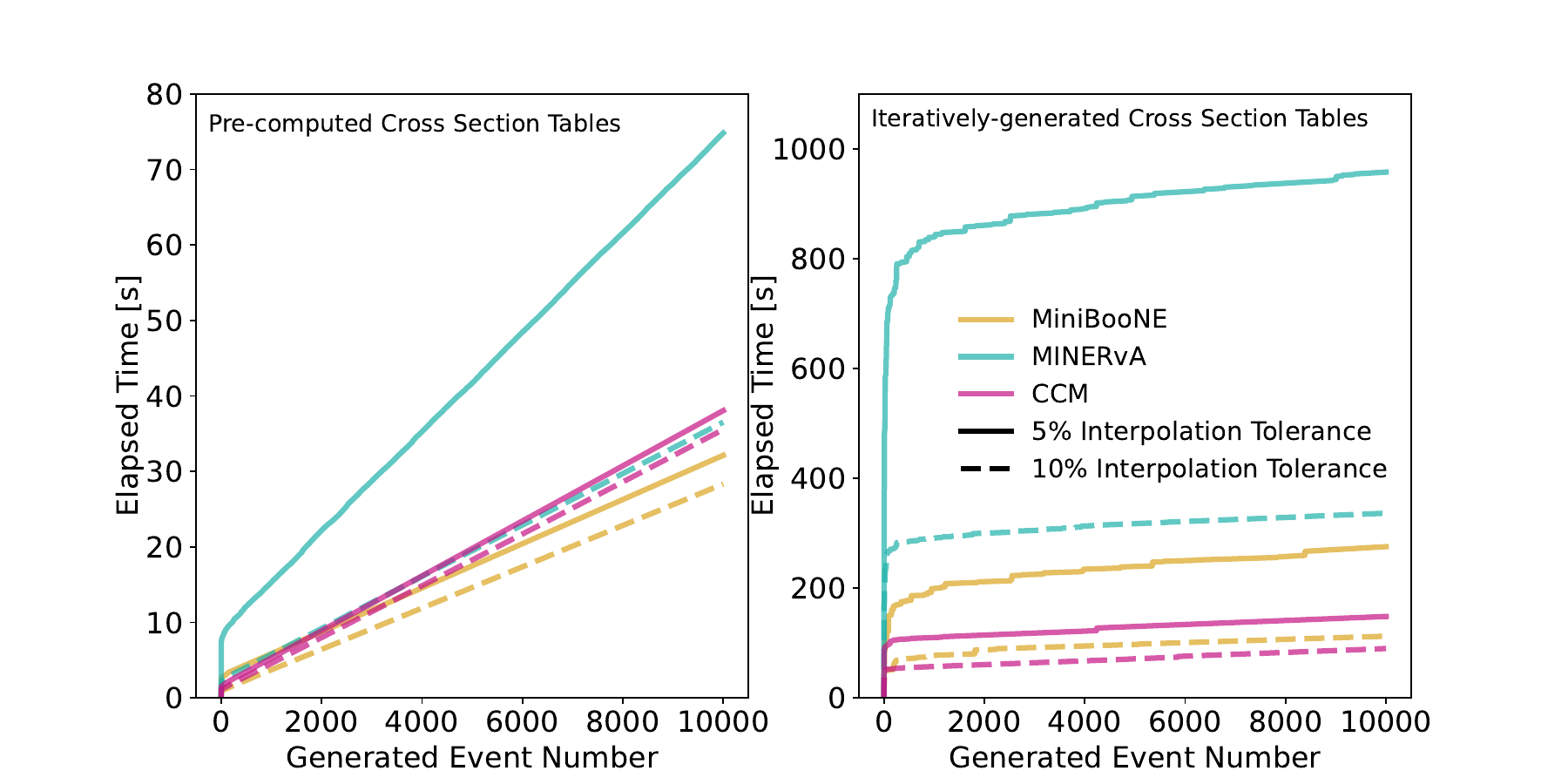}
    \caption{Elapsed time as a function of generated event number for the examples presented in \cref{sec:dipole_examples}. The left (right) subfigure reflects the case in which cross section tables are computed before (during) event generation. Two different interpolation tolerances are shown, which influence the computational efficiency of the generation.}
    \label{fig:dipole_efficiency}
\end{figure*}

\Cref{fig:dipole_efficiency} shows the elapsed time to generate $10,000$ events in the dipole-portal HNL examples presented in \cref{sec:dipole_examples}.
We report efficiencies separately for the case in which cross section tables are computed ahead of time and computed iteratively during generation, as discussed in \cref{sec:dark_news}.
One can see that in the latter case, the tables are mostly filled during the first $\mathcal{O}(1000)$ generated events, and the tables are most expensive to compute in the \minerva example.
Even in the pre-computed case, one can see that \siren takes a non-negligible time to load the cross section tables, especially in the \minerva case.
We also consider two different interpolation tolerances: 5\% and 10\%.
The simulation is generally faster in the 10\% case, as table queries are less computationally expensive.

\begin{table*}[]
    \centering
    \begin{tabular}{c|c|c}
        Simulation case & Generation time per event [s] & Weight calculation time per event [s]  \\
        \hline
        $\nu_\mu$ DIS in IceCube & $7.37^{+ 1.24}_{- 1.31} \times 10^{-5}$ & $12.83^{+ 1.45}_{- 3.36} \times 10^{-5}$ \\[1mm]
        $\nu_\mu$ DIS in DUNE & $5.63^{+ 0.98}_{- 0.76} \times 10^{-5}$ & $8.63^{+ 1.41}_{- 1.93} \times 10^{-5}$ \\[1mm]
        $\nu_\mu$ DIS in ATLAS & $3.74^{+ 0.14}_{- 0.10} \times 10^{-5}$ & $6.58^{+ 0.21}_{- 0.29} \times 10^{-5}$ \\[1mm]
        Dipole-portal HNLs in MiniBooNE & $2.97^{+ 0.04}_{- 0.07} \times 10^{-3}$ & $2.07^{+ 0.03}_{- 0.25} \times 10^{-3}$ \\[1mm] 
        Dipole-portal HNLs in \minerva & $4.72^{+ 5.93}_{- 1.12} \times 10^{-3}$ & $4.00^{+ 1.91}_{- 0.42} \times 10^{-3}$ \\[1mm] 
        Dipole-portal HNLs in CCM & $3.83^{+ 0.05}_{- 0.07} \times 10^{-3}$ & $4.25^{+ 0.08}_{- 0.13} \times 10^{-3}$\\[1mm] 
    \end{tabular}
    \caption{Computational efficiency of each of the examples discussed in \cref{sec:examples}, represented by the event generation and weight calculation time per event. The table entries reflect the median and $\pm 1\sigma$ width of the distributions shown in \cref{app:comp_efficiency}.}
    \label{tab:computation_efficiency}
\end{table*}

%% file: sections/05_weighting.tex
\section{Calculating Event Weights} \label{sec:weighting}

The weight calculation in \siren is inspired by that of \prevleptoninjector~\cite{IceCube:2020tcq}, but has been extended to support multiple interactions for a given simulation event.
We also support interaction probabilities with non-trivial dependence on the neutrino path, i.e. beyond the DIS-based assumption that interaction density is proportional to column density.
\siren is designed to be reweightable, meaning that the user can use the same simulation set to reflect different physical scenarios by recomputing event weights of the simulation set.
This is very useful in cases where \siren is used as the injection and weighting interface for a more computationally expensive detector simulation.
This feature can also enable efficient scans of the parameter space of a new physics model, assuming the same simulation set can be reweighted to evaluate event rates for different values of the model parameters.

Weight calculation in \siren is handled by the \bcname{Weighter} class, which works as follows.
Suppose we have generated a set of $N_{\rm gen}$ events in \siren involving a single process, either scattering or decay.
The physical weight $w$ of each event is generically given by
\begin{equation} \label{eq:weight}
w = \frac{p_{\rm phys}}{N_{\rm gen} p_{\rm gen}},
\end{equation}
where $p_{\rm gen}$ and $p_{\rm phys}$ are the generation and physical probability of the event, respectively.
This can be thought of as a two-step process, in which the denominator of \cref{eq:weight} first transforms events into a uniform distribution in all generated variables, and then the numerator transforms events into the desired physical distribution. 

The generation probability involves a product over the sampling probability of each of the generation distributions of \cref{fig:distributions_hierarchy} as well as terms related to the interaction model, which typically looks like
\begin{equation} \label{eq:gen_prob}
\begin{split}
p_{\rm gen} = &p_{\rm gen}^{\rm interaction~type} p_{\rm gen}^{\rm kinematics} \\
\times &p_{\rm gen}^{\rm energy} p_{\rm gen}^{\rm direction} p_{\rm gen}^{\rm helicity} p_{\rm gen}^{\rm mass} p_{\rm gen}^{\rm vertex}.
\end{split}
\end{equation}
The probability of undergoing the generated interaction type is given by
\begin{equation} \label{eq:gen_int_type_prob}
p_{\rm gen}^{\rm interaction~type} = \frac{n_t \sigma^{t,i}_{\rm gen}~\textbf{or}~(L^{d}_{\rm gen})^{-1}}{\sum_d (L^{d}_{\rm gen})^{-1} + \sum_{t,i} n_t \sigma^{t,i}_{\rm gen}}
\end{equation}
where $n_t$ is the number density of the struck target particle $t$, $\sigma^{t,i}_{\rm gen}$ is the generation-level cross section of the interaction $i$ on the target $t$, and $L^{d}_{\rm gen}$ is the generation-level lab frame decay length of the decay process $d$.
The first (second) term in the numerator corresponds to the case in which the generated interaction is a scattering (decay) process.
The probability of producing the generated final state kinematics is given by
\begin{equation} \label{eq:gen_kin_prob}
p_{\rm gen}^{\rm kinematics} = \frac{1}{\sigma^{t,i}_{\rm gen}} \frac{\partial \sigma^{t,i}_{\rm gen}}{\partial {\Pi}}~\textbf{or}~\frac{1}{\Gamma^{d}_{\rm gen}} \frac{\partial \Gamma^{d}_{\rm gen}}{\partial {\Pi}},
\end{equation}
where $\Pi$ is the phase space of the differential cross section or decay width $\Gamma^d_{\rm gen}$.
Again, the first (second) term corresponds to the case in which the generated interaction is a scattering (decay) process.
The exact forms of $p_{\rm gen}^{\rm energy}$, $p_{\rm gen}^{\rm direction}$ and $p_{\rm gen}^{\rm vertex}$ depend on the specified energy directional and vertex generation distributions.
Finally, $p_{\rm gen}^{\rm helicity}$ and $p_{\rm gen}^{\rm mass}$ evaluate to zero or one depending on whether the primary particle under consideration has a helicity and mass consistent the possible generated values.

The physical probability involves a product over user-specified physical distributions as well as terms related to the interaction model, overall interaction probability, and interaction location, which typically looks like
\begin{equation} \label{eq:phys_prob}
\begin{split}
p_{\rm phys} = & A_{\rm phys} p_{\rm phys}^{\rm interaction~type} p_{\rm phys}^{\rm kinematics} p_{\rm phys}^{\rm interaction} \\
\times &p_{\rm phys}^{\rm energy} p_{\rm phys}^{\rm direction} p_{\rm phys}^{\rm helicity} p_{\rm phys}^{\rm mass} p_{\rm phys}^{\rm vertex},
\end{split}
\end{equation}
where $A_{\rm phys}$ is an overall factor reflecting the physical normalization of any of the user-specified physical distributions.
In practice, this is used to specify the normalization of the incoming neutrino flux, either as the integral of a \cname{PrimaryEnergyDistribution} or as a separate normalization constant.
The terms related to the interaction model in \cref{eq:phys_prob}, $p_{\rm phys}^{\rm interaction~type}$ and $p_{\rm phys}^{\rm kinematics}$, follow from \cref{eq:gen_int_type_prob,eq:gen_kin_prob} considering the physical scattering cross sections $\sigma^{t,i}_{\rm phys}$ and decay lengths $L^d_{\rm phys}$.
The physical probability that an incoming particle interacted within the injection bounds is given by
\begin{equation}
p_{\rm phys}^{\rm interaction} = 1 - \exp \bigg[ -\int_{\ell_i}^{\ell_f} d \ell \bigg(L_{\rm phys}^{-1} + \sum_{t,i} n^t (\ell) \sigma_{\rm phys}^{t,i} \bigg) \bigg],
\end{equation}
where $L_{\rm phys}$ is the total physical decay length of the particle considering all possible decay interactions, and $\ell_i$ and $\ell_f$ are the injection bounds.
As in \cref{eq:gen_prob}, the energy, direction, helicity, and mass terms in \cref{eq:phys_prob} depend on the user-specified physical distributions.
The $p_{\rm phys}^{\rm vertex}$ term is special and is always calculated using the physical interaction depth along the particle's path,
\begin{equation}
p_{\rm phys}^{\rm vertex} = \frac{\exp\big[  -\int_{\ell_i}^\ell d\ell \big( L_{\rm phys}^{-1} + \sum_{t,i} n^t (\ell) \sigma_{\rm phys}^{t,i} \big) \big]}{\int_{\ell_i}^{\ell_f} d \ell \exp\big[  -\int_{\ell_i}^{\ell} d\ell \big( L_{\rm phys}^{-1} + \sum_{t,i} n^t (\ell) \sigma_{\rm phys}^{t,i} \big) \big]}
\end{equation}

The calculation outlined above can be extended to a generation case involving secondary processes beyond the initial primary process.
The weight of each event in this case looks like
\begin{equation}
w = \frac{1}{N_{\rm gen}} \prod_{{\rm processes}~p} \frac{p_{\rm phys}^p}{p_{\rm gen}^p}.
\end{equation}
The only difference here is that for secondary processes, the properties of the primary particle are already fixed.
This means that the energy, direction, helicity, and mass distributions do not show up in \cref{eq:gen_prob,eq:phys_prob}.
Further, the generation-level vertex distribution for secondary processes includes the option to require interactions to happen within a user-specified fiducial volume.

Finally, \siren supports the calculation of weights for events generated by multiple injectors, as long as they have the same set of available processes.
In this case, the weights are given by
\begin{equation}
w = \bigg[ \sum_{{\rm injectors}~j} N_{\rm gen}^j \prod_{{\rm processes}~p} \frac{p_{\rm gen}^{p,j}}{p_{\rm phys}^{p}}\bigg]^{-1}.
\end{equation}

%% file: sections/06_examples.tex
\section{Examples}
\label{sec:examples}

This section introduces two sets of examples demonstrating the simulation of different physics models using \siren.
The first set considers deep inelastic scattering of muon neutrinos in IceCube, DUNE, and ATLAS, and the second set considers the production and decay of heavy neutral leptons via a transition magnetic moment in MiniBooNE, \minerva, and CCM.
These examples are meant to serve as templates from which users can develop their own simulation scripts.
The Python-based simulation script for each example is available in the \texttt{Resources/Examples} directory of the repository.
Two examples are provided in \cref{app:code_examples}.

\subsection{Example 1: $\nu_\mu$ DIS in IceCube, DUNE, and ATLAS}
\label{sec:DIS_examples}

\begin{figure*}[h]
    \centering
    \includegraphics[width=0.45\textwidth]{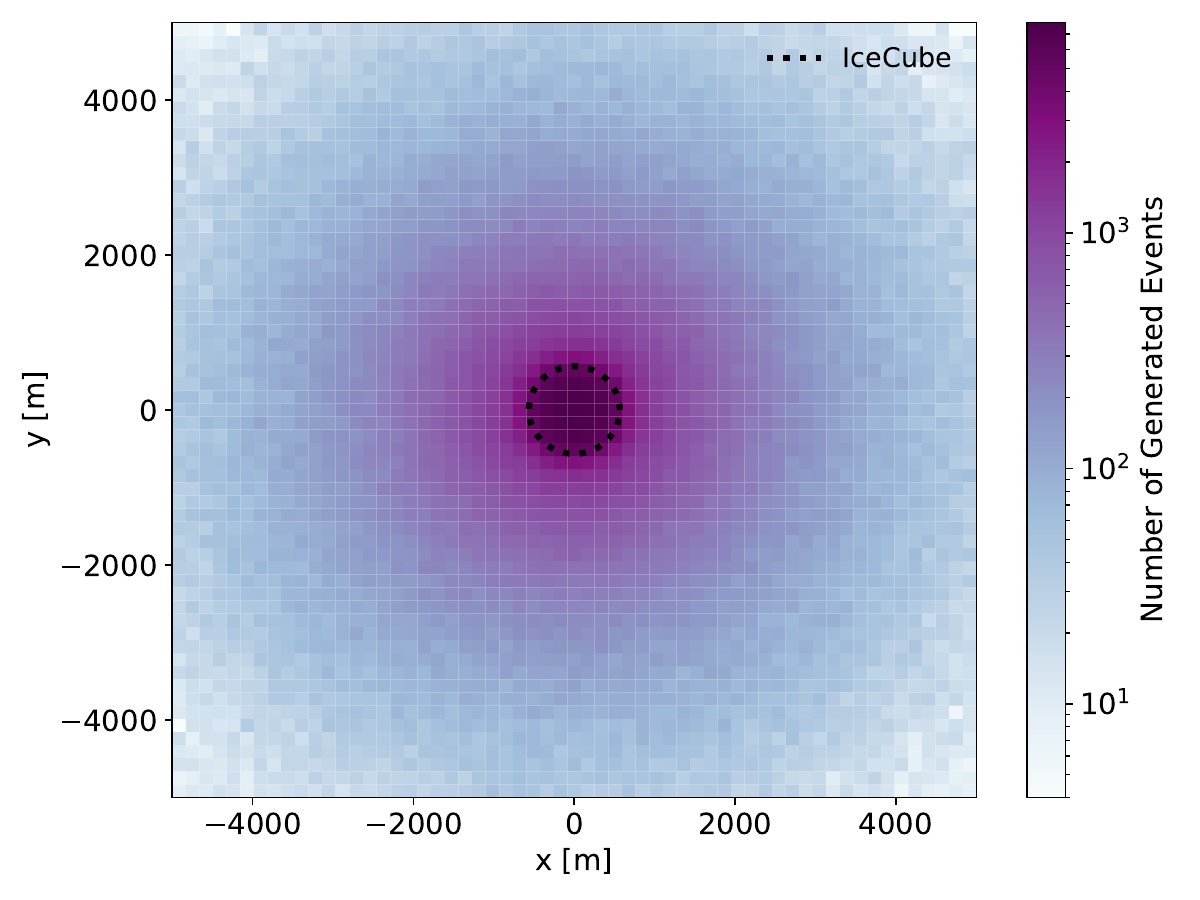}
    \includegraphics[width=0.45\textwidth]{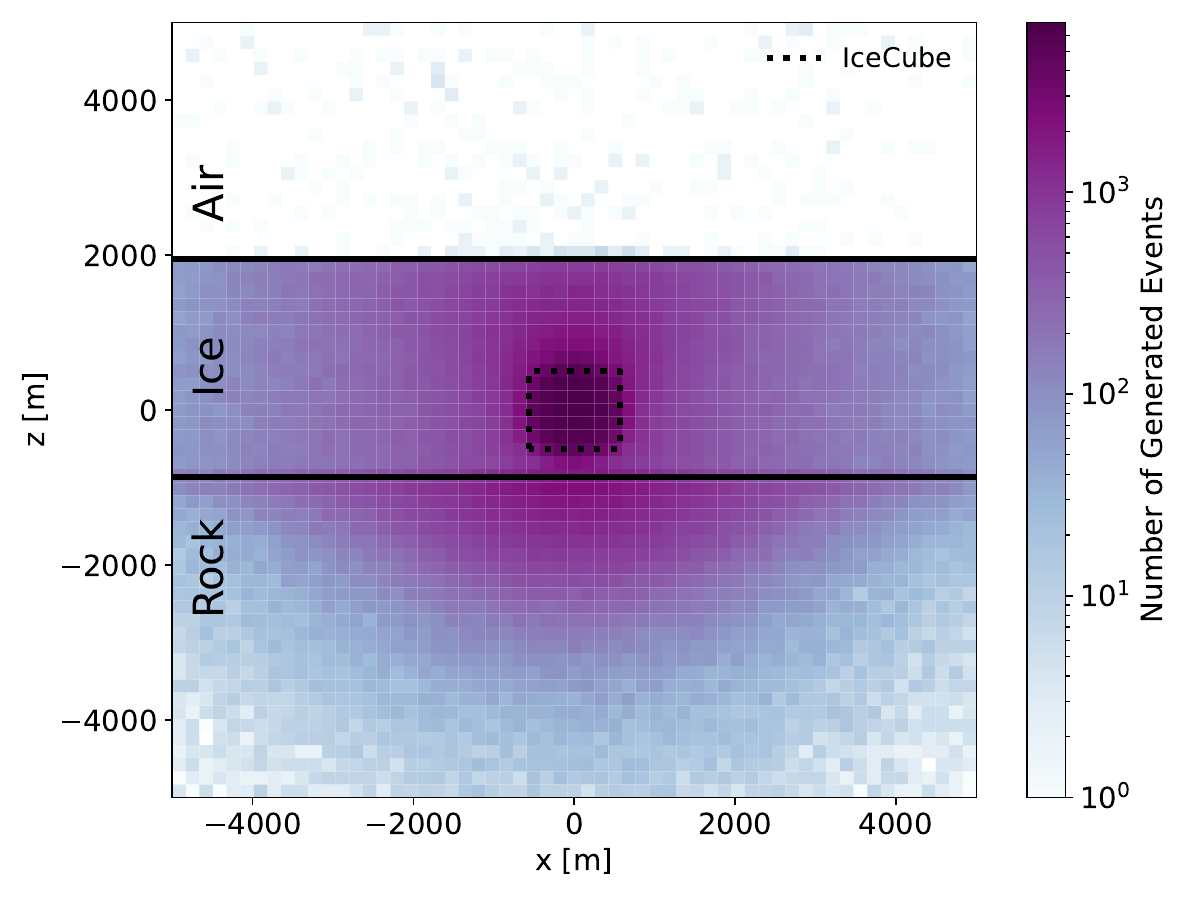}
    \\
    \includegraphics[width=0.45\textwidth]{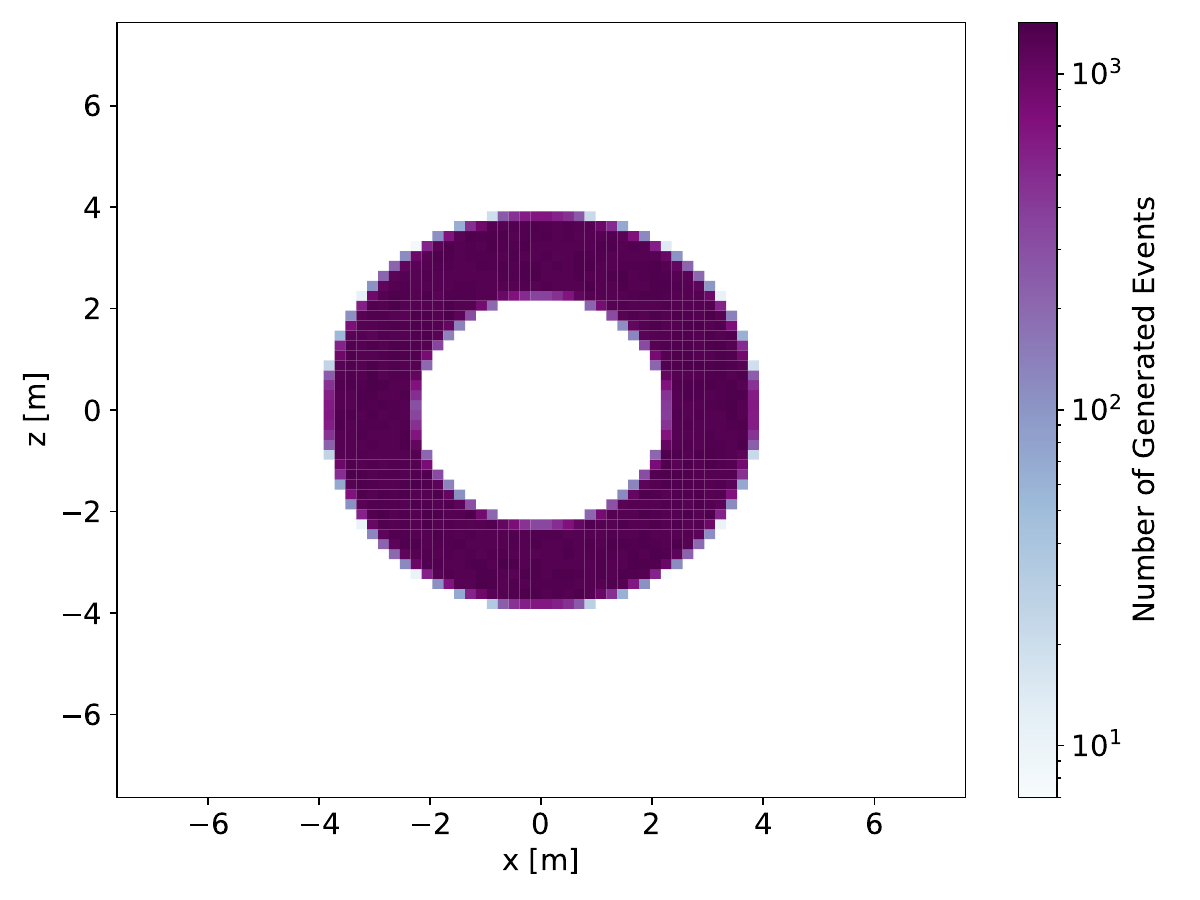}
    \includegraphics[width=0.45\textwidth]{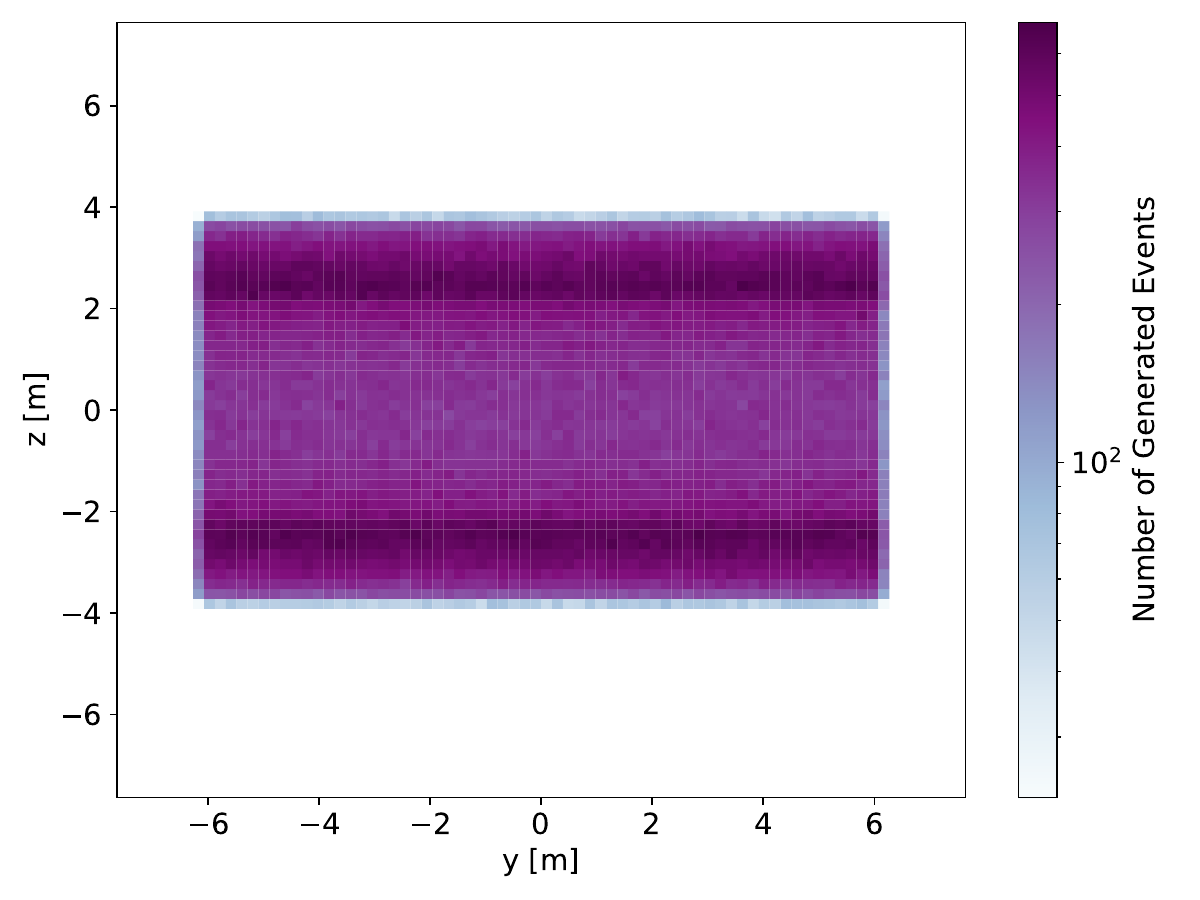}
    \caption{Distributions of generated $\nu$DIS interaction locations for ranged and volume injection in IceCube (top) and the ATLAS hadronic calorimeter (bottom), respectively.}
    \label{fig:DIS_locations}
\end{figure*}

\begin{figure}[h]
    \centering
    \includegraphics[width=\linewidth]{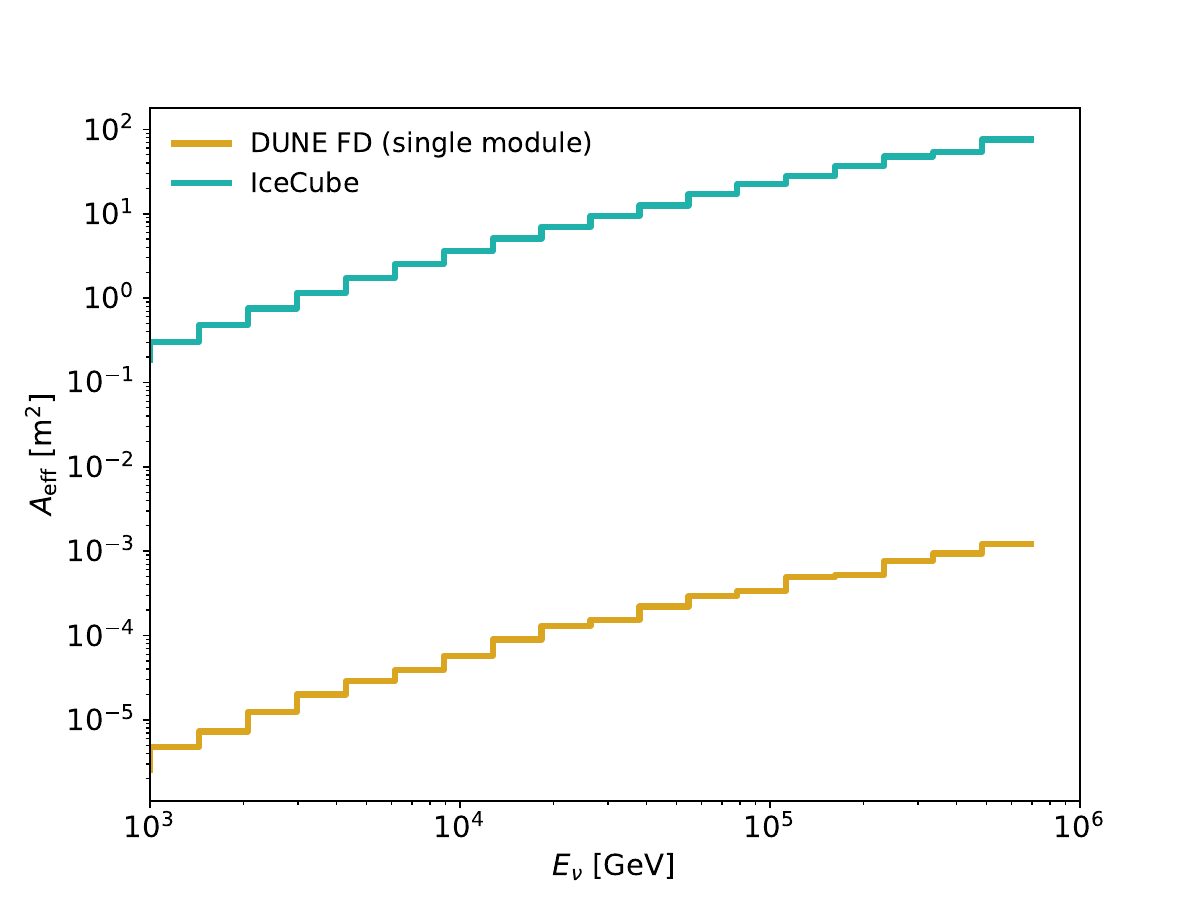}
    \caption{The effective area of a single DUNE far detector module and the IceCube detector as a function of initial neutrino energy. The effective area is calculated according to \cref{eq:Aeff}.}
    \label{fig:Aeff}
\end{figure}

\begin{figure}[h]
    \centering
    \includegraphics[width=\linewidth]{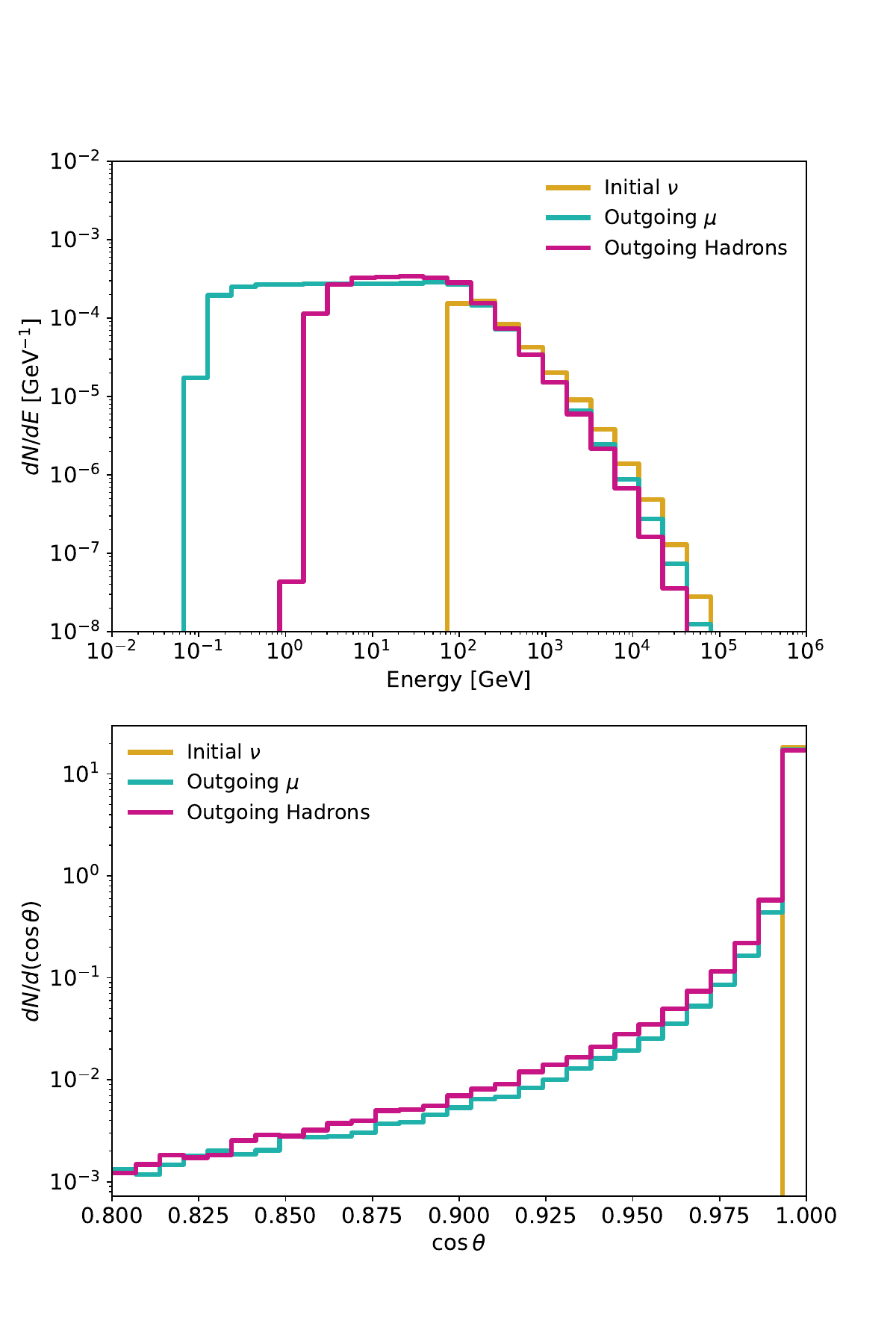}
    \caption{Physically-weighted energy (top) and angular (bottom) distributions of the initial neutrino and outgoing muon and hadronic system, for the ATLAS volume injection example, for the case of muon charged-current events. The energy distribution shows the shape of the injected neutrino energy spectrum convolved with the cross section. The reaction products, muons and hadronic showers, cascade to lower energies as expected. The angular distribution shows the relatively small opening angle associated with the reaction products compared to the initial neutrino direction, as characteristic of high-energy neutrino interactions.}
    \label{fig:ATLAS_kinematics}
\end{figure}

The first set of examples simulates the deep inelastic scattering (DIS) of muon neutrinos within and/or surrounding the IceCube, DUNE, and ATLAS detectors.
The IceCube example is meant to emulate the original use case of the \prevleptoninjector software package~\cite{IceCube:2020tcq}.
\Cref{app:LI_validation} goes into more detail on the backward compatibility of \siren regarding the \prevleptoninjector use cases.
The DUNE example follows from Ref.~\cite{Schneider:2021wzs}, which made the first extensions of \prevleptoninjector to compute DUNE's sensitivity to atmospheric neutrinos.
The ATLAS example follows from Ref.~\cite{Wen:2023ijf}, which made further extensions to \prevleptoninjector to compute ATLAS's sensitivity to the predicted high energy neutrino flux from supernovae ~\cite{Murase:2017pfe}.

We begin with the IceCube and DUNE examples, which have a similar configuration.
Both simulations are concerned mainly with the rate of through-going muons from $\nu_\mu$ DIS interactions in the surrounding material.
Therefore, we use the ranged injection scheme described in \cref{sec:injection_methodologies}, where the range is taken to be the length for which 99.9\% of muons of a given energy will survive to the detector~\cite{Chirkin:2004hz}.
The initial neutrino energy is sampled from a power law with a spectral index of 2, bounded between $1\;{\rm TeV}$ and $1\;{\rm PeV}$.
The direction is sampled isotropically.
The location of the neutrino interaction is sampled from the column depth position distribution as described in \cref{sec:injection_methodologies}, using the muon survival length to determine the range, which depends on the muon (and thus neutrino) energy.
For DIS, sampling in column depth is equivalent to sampling according to the interaction length.
This scheme is intended to enable the efficient simulation of through-going muons in IceCube and DUNE.
The disk radius and endcap length for ranged injection are chosen to encompass each detector; in IceCube (DUNE), both the radius and endcap length are set to 600 (60)~m.
The IceCube detector is approximated by a cylinder of ice with a radius of 546.2~m and height of 1000~m, corresponding to a volume of 1~${\rm km}^{3}$.
For DUNE, we consider a single far detector module approximated by a $14\;{\rm m} \times 58.2\;{\rm m} \times 12\;{\rm m}$ rectangular prism of liquid argon.

Once the properties of the initial neutrino are fixed and an interaction location is sampled, the properties of the outgoing muon and hadronic system are determined from the differential cross section.
Two kinematic variables are necessary to specify particle kinematics in DIS interactions.
As described in \cref{sec:provided_xsec}, we use the Metropolis-Hasting algorithm~\cite{Metropolis:1953am,Hastings:1970aa} to sample Bjorken x and y from the double-differential charged-current DIS cross section computed in Ref.~\cite{Cooper-Sarkar:2011jtt}.
The physical weight of each event is computed according to the procedure outlined in \cref{sec:weighting}.
The final result of this scheme is a set of events, each represented by an \cname{InteractionTree}, with corresponding physical weights computed according to the procedure outlined in \cref{sec:weighting}.
Weighted distributions computed using these events represent probability densities for a single incident neutrino, incorporating the interaction probability of that neutrino.
One can use this to compute event rate distributions for different incident neutrino rates reflecting different flux models.

The ATLAS example is configured differently.
ATLAS \cite{ATLAS:2008xda} is a large multipurpose collider detector located at CERN. While primarily designed to study proton-proton collisions of the LHC beam, it has some features that make it viable for direct neutrino detection.
The hadronic calorimeter \cite{ATLAS:1996aa} is an approximately $4 \: \rm kT$ barrel-shaped ATLAS sub-detector made primarily of metal and plastic scintillator plates. 
The mass of this detector as well as its coarse segmentation makes the detection of neutrino-indiced hadronic showers possible. 
This example is based on the study presented in Ref.~\cite{Wen:2023ijf}, which used an earlier iteration of \siren to compute all-flavor $\nu$ DIS interaction rates within the ATLAS hadronic calorimeter from high-energy supernova neutrinos.
Supernova explosions may produce a high-energy flux of neutrinos above $100\:\rm GeV$, depending on the type of supernova and its distance to the detector \cite{Murase:2017pfe}.
The hadronic calorimeter is modelled as a uniform-density cylinder of iron with length $12\: \rm m$, inner radius $2.3 \: \rm m$, and outer radius $3.8\: \rm m$, approximating the dimensions \cite{ATLAS:1996aa} of the hadronic calorimeter.
To find the number of neutrino interactions that take place inside the hadronic calorimeter, we perform a volume injection described in \cref{sec:injection_methodologies} (in contrast to a ranged injection) for this detector geometry.
The primary interaction vertex is chosen randomly from a position within the volume and the associated neutrino properties are weighted according to given flux and cross section distributions.
The neutrino flux from Ref.~\cite{Murase:2017pfe} is sampled with an inverse-CDF method from a tabulated distribution (since it is not a simple power law) and we sample the same neutrino-nucleus DIS cross sections used also for the other examples.
As we are considering a supernova source, we only consider neutrinos from a single direction.

In \cref{fig:DIS_locations}, we show the generation-level distributions of interaction locations for the IceCube and ATLAS examples.
The IceCube figures demonstrate the utility of ranged injection--most of the simulated neutrino interactions occur outside the detector volume.
Moreover, since locations are sampled according to column depth along the neutrino line-of-sight, one can see clear differences between the regions of the detector environment consisting of rock, ice, and air.
The DUNE interaction location distribution is largely similar to the IceCube case, with the only differences coming from the smaller disk radius and endcap length as well as the different detector environments.
The ATLAS figures demonstrate the behavior of volume injection--one can clearly see the shape of the hadronic calorimeter, throughout which neutrino interactions are generated uniformly.

Using the output from the IceCube and DUNE examples, we can compute the effective area of each detector for through-going muons.
The effective area describes the representative area of a detector that can observe 100\% of the incident through-going muon neutrino interactions.
This incorporates effects from the detector size and neutrino interaction probability.
It can also be defined to capture the detector selection efficiency; however, we ignore this effect here.
The effective area $A_{\rm eff}$ is defined mathematically by the relation
\begin{equation}
\dot{N}_\nu = \int d\Omega \int  d E_\nu A_{\rm eff}(E_\nu,\Omega) \phi(E_\nu,\Omega),
\end{equation}
where $\dot{N}_\nu$ is the observed neutrino event rate in units of $[\nu\;{\rm s}^{-1}]$, $\phi$ is the neutrino flux in units of $[\nu\;{\rm GeV}^{-1}\;{\rm m}^{-2}\;{\rm sr}^{-1}\;{\rm s}^{-1}]$, and $E_\nu$ is the energy of the neutrino.
The solid angle-integrated effective area is given by
\begin{equation}\label{eq:Aeff}
\begin{split} 
\hat{A}_{\rm eff}(E_\nu) &\equiv \frac{1}{\int d\Omega \phi(E_\nu, \Omega)} \int d\Omega A_{\rm eff}(E_\nu,\Omega)\phi(E_\nu, \Omega) \\
&= \frac{d \dot{N}_\nu}{d E_\nu} \frac{1}{\int d\Omega\phi(E_\nu, \Omega)} = \frac{d \dot{N}_\nu}{d E_\nu} \frac{4 \pi}{\phi(E_\nu)},
\end{split}
\end{equation}
where $\phi(E_\nu) \equiv \phi(E_\nu,\Omega) / 4\pi$ assuming an isotropic flux.
The average effective area across an energy bin $i$ is computed using the \siren weights $w_j$ of events in that bin by
\begin{equation} \label{eq:Aeff_discrete}
\begin{split}
    \hat{A}_{\rm eff}^i &= \frac{1}{E_\nu^{i+1} - E_\nu^i}\int_{E_\nu^i}^{E_\nu^{i+1}} d E_\nu \frac{d\dot{N}_\nu}{d E_\nu} \frac{4\pi}{\phi(E_\nu)} \\
    &= \frac{4\pi}{E_\nu^{i+1} - E_\nu^i} \sum_j \frac{w_j}{\phi(E_\nu^j)},
\end{split}
\end{equation}
where $E_\nu^j$ is the neutrino energy event of event $j$.
For this calculation of the IceCube and DUNE effective areas, we restrict ourselves to observable events, i.e. events for which the muon passes through the active volume of each detector.
This is possible through the \cname{DetectorModel} interface of \siren.
\Cref{fig:Aeff} shows the effective area for the entire IceCube detector and a single DUNE far detector module computed according to \cref{eq:Aeff_discrete}.
The ratio between the effective area of each experiment reflects the relative volume of the IceCube ($\sim 1\;{\rm GT}$) and DUNE ($\sim 10\;{\rm kT}$) detectors.

In \cref{fig:ATLAS_kinematics}, we show the energy and angular distributions of the particles involved in the $\nu_\mu$ CC DIS interactions inside the ATLAS hadronic calorimeter.
Since the neutrino flux here comes from a distant supernova, they arrive from a single direction in the detector.
These distributions can be used in conjunction with information about ATLAS's detector response to directly determine the detector's sensitivity to high-energy supernova neutrinos, as demonstrated in Ref.~\cite{Wen:2023ijf}.
Similar simulations can be performed for through-going muons from $\nu_\mu$ CC DIS interactions in the bedrock surrounding ATLAS; this has also been demonstrated in Ref.~\cite{Wen:2023ijf}.
In this case, one can simulate the muon propagation using a tool such as \cname{PROPOSAL}~\cite{Koehne:2013gpa} to determine the survival probability and modulated energy distribution.

\subsection{Heavy Neutral Leptons in MiniBooNE, MINERvA and CCM}
\label{sec:dipole_examples}

\begin{figure*}[h]
    \centering
    \includegraphics[width=0.3\textwidth]{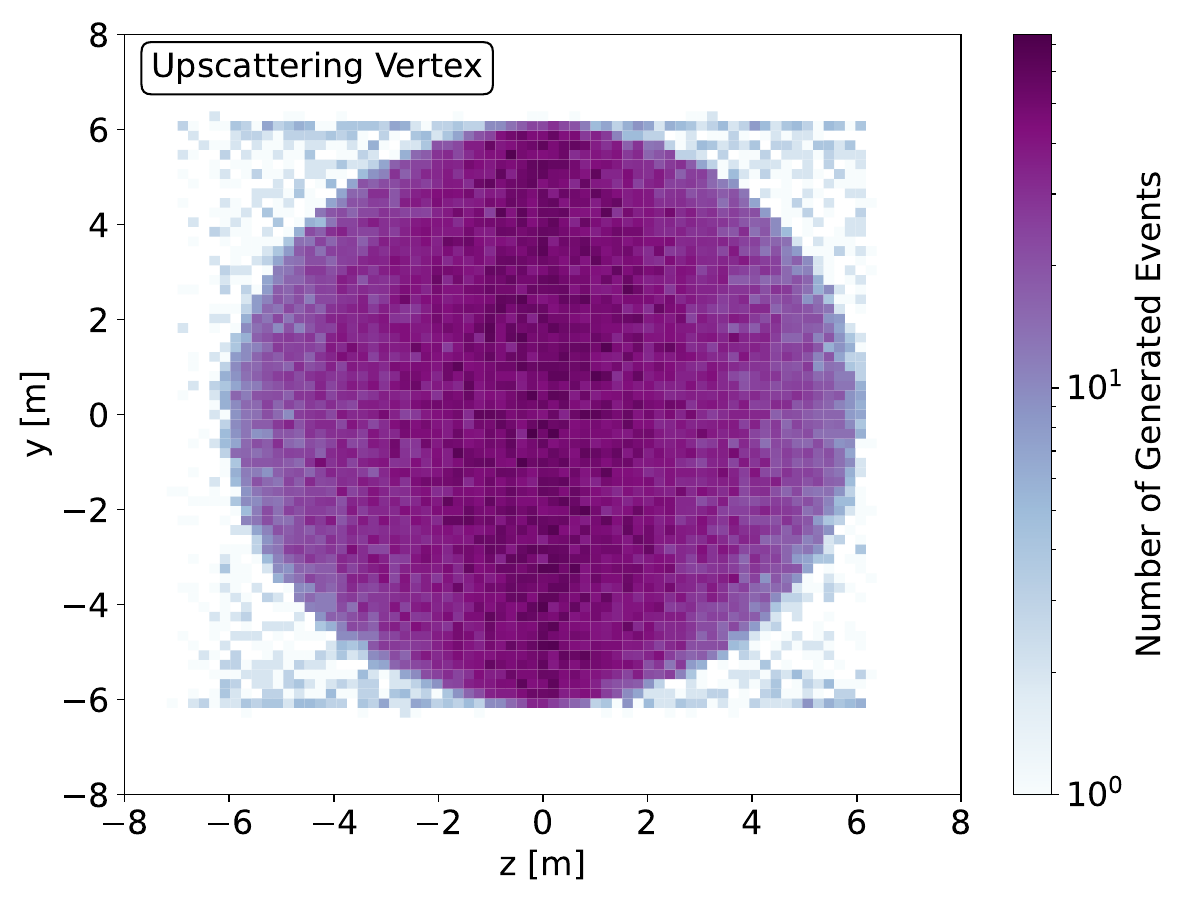}
    \includegraphics[width=0.3\textwidth]{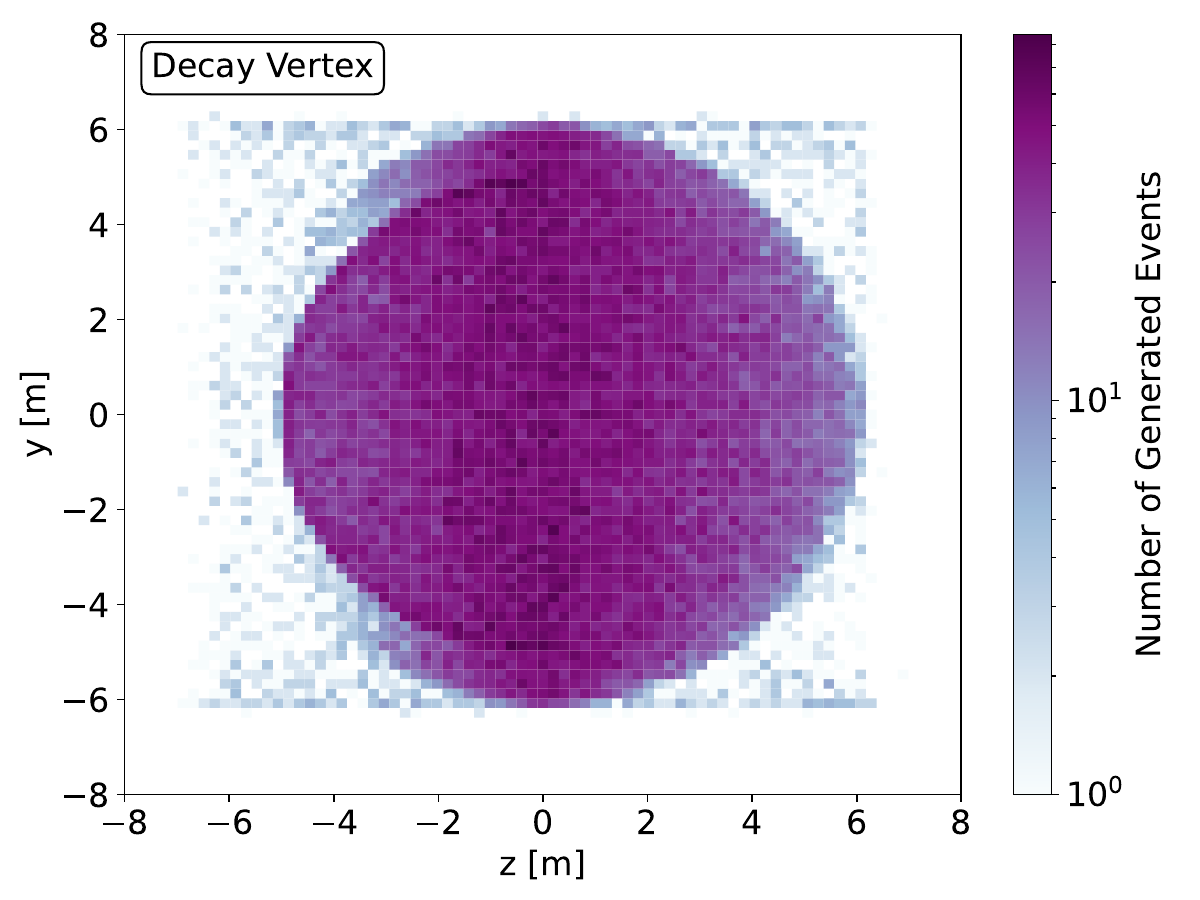}
    \includegraphics[width=0.3\textwidth]{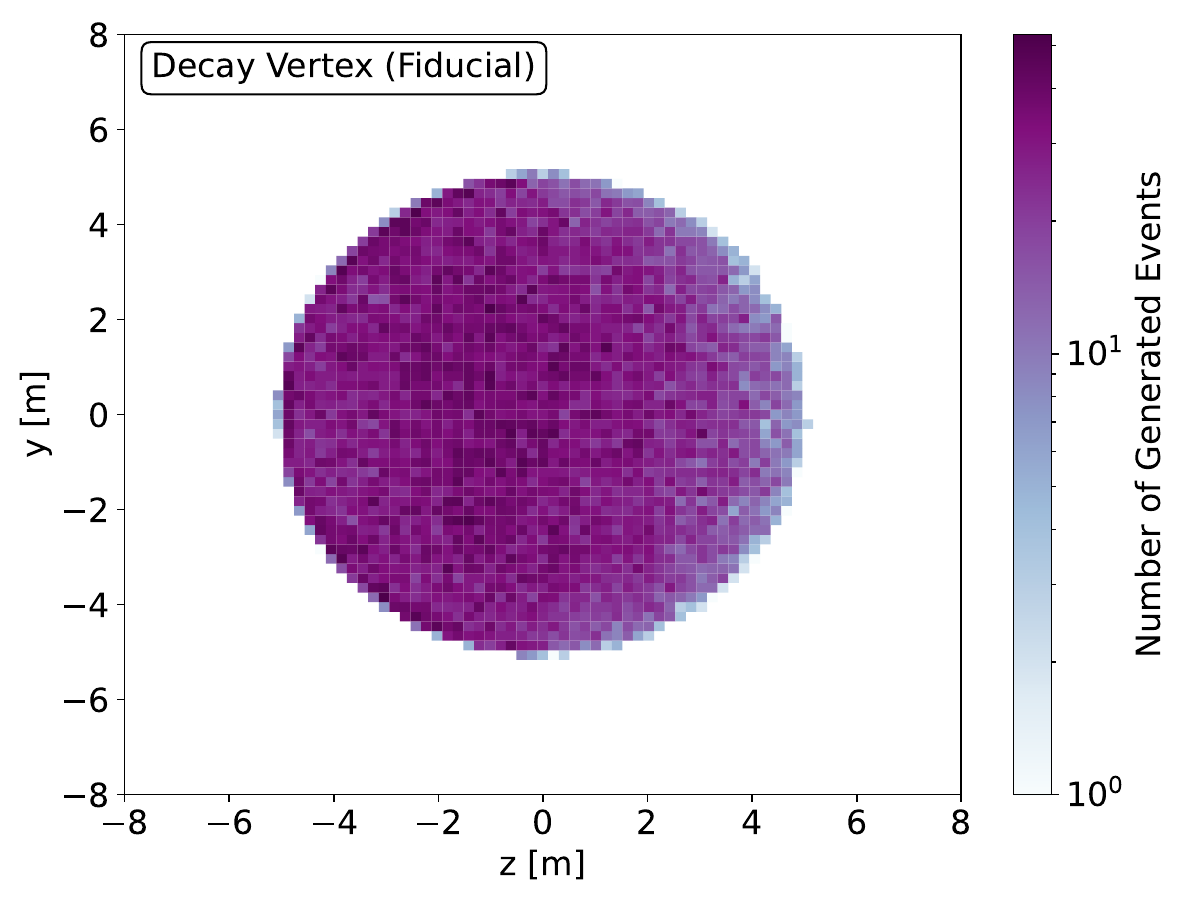}
    \includegraphics[width=0.3\textwidth]{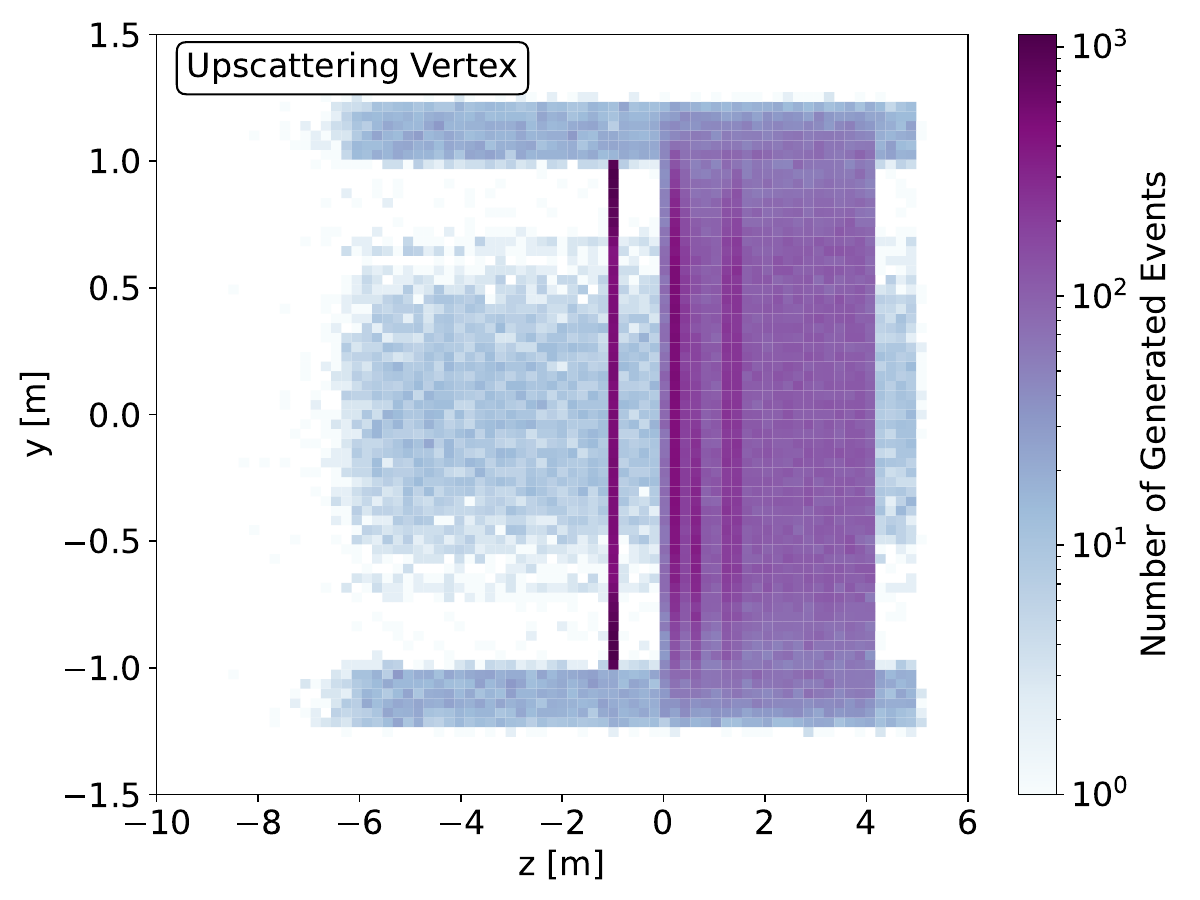}
    \includegraphics[width=0.3\textwidth]{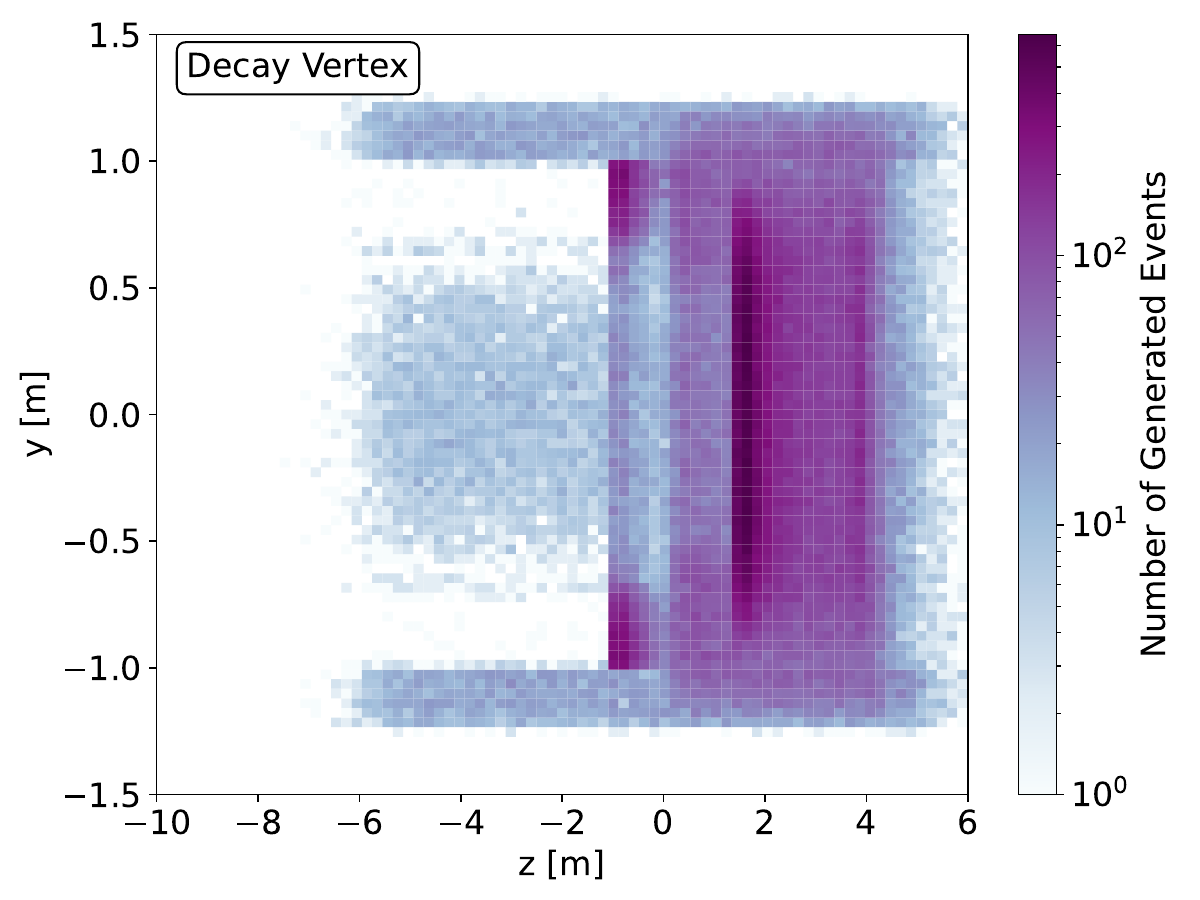}
    \includegraphics[width=0.3\textwidth]{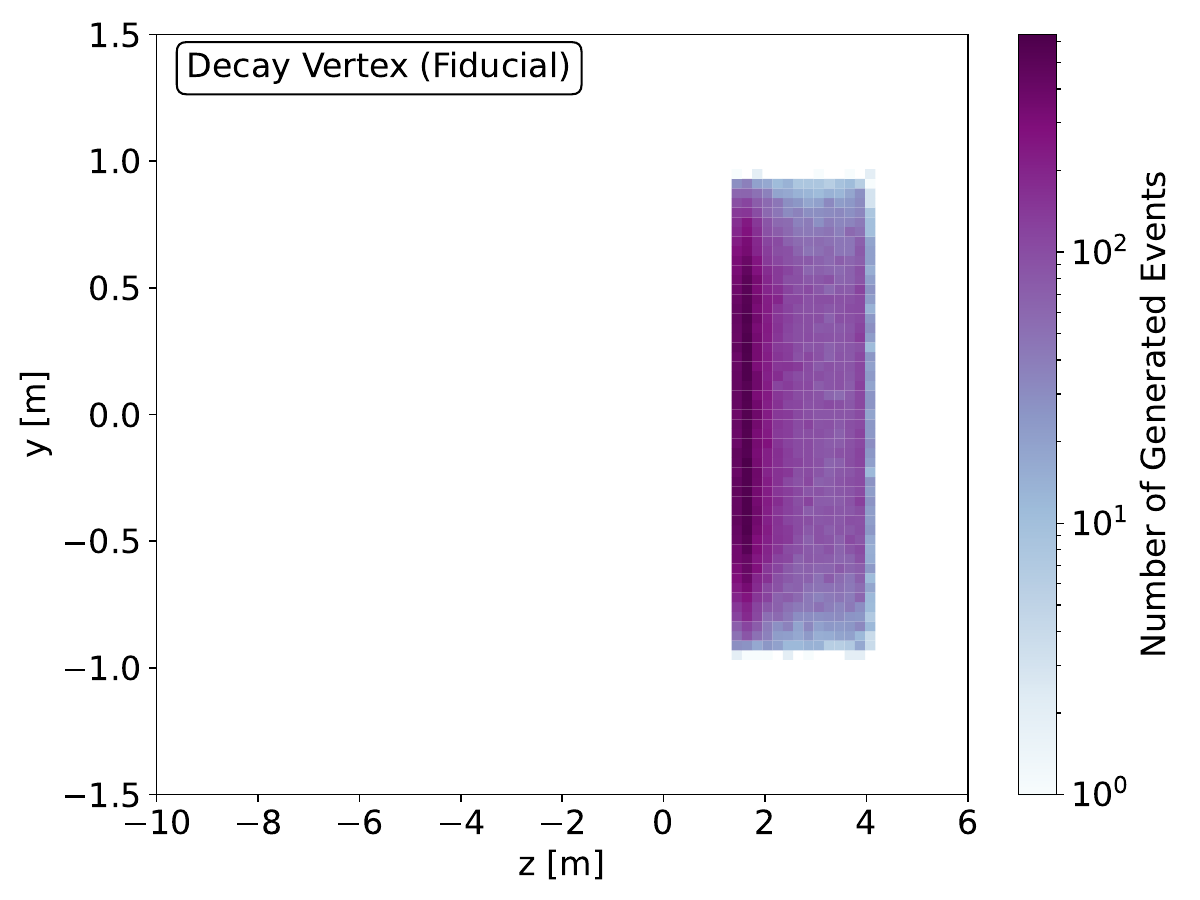}
    \includegraphics[width=0.3\textwidth]{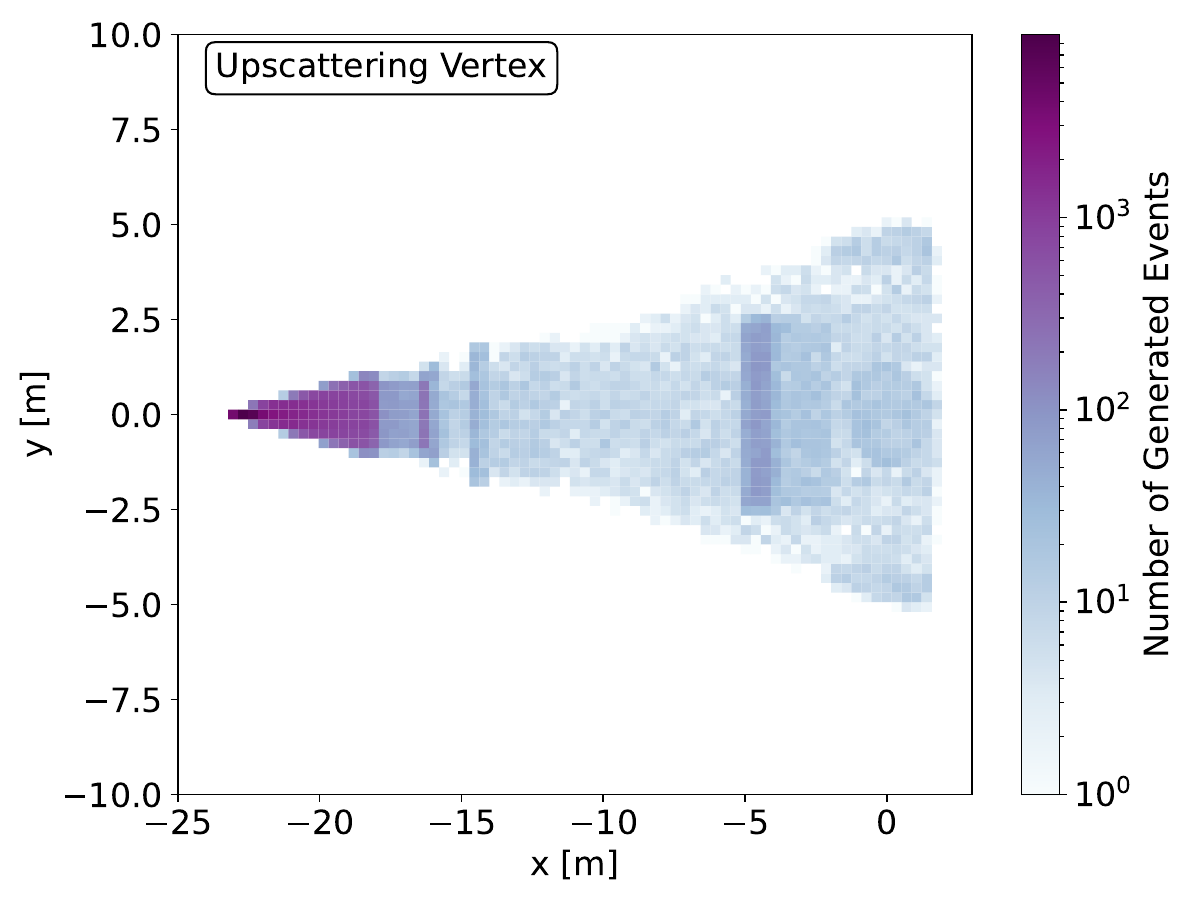}
    \includegraphics[width=0.3\textwidth]{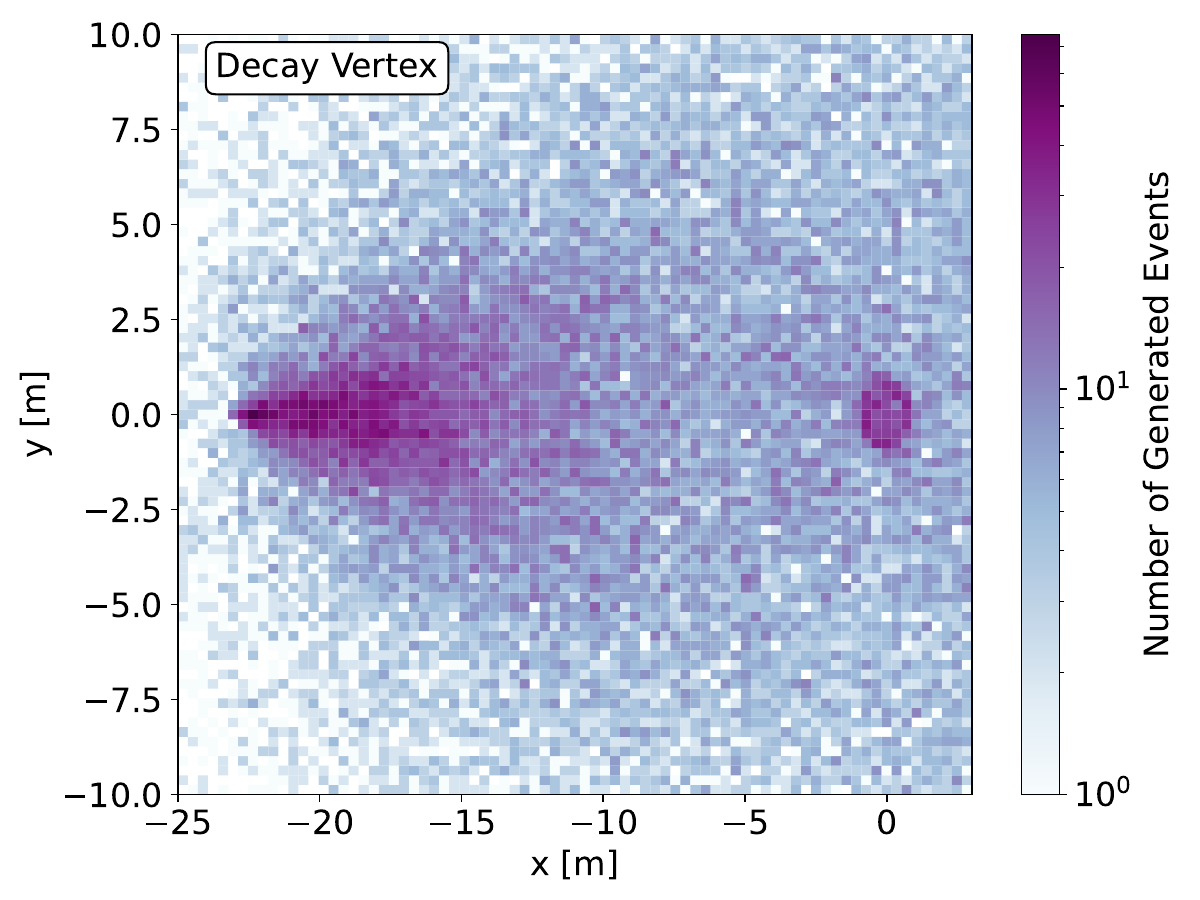}
    \includegraphics[width=0.3\textwidth]{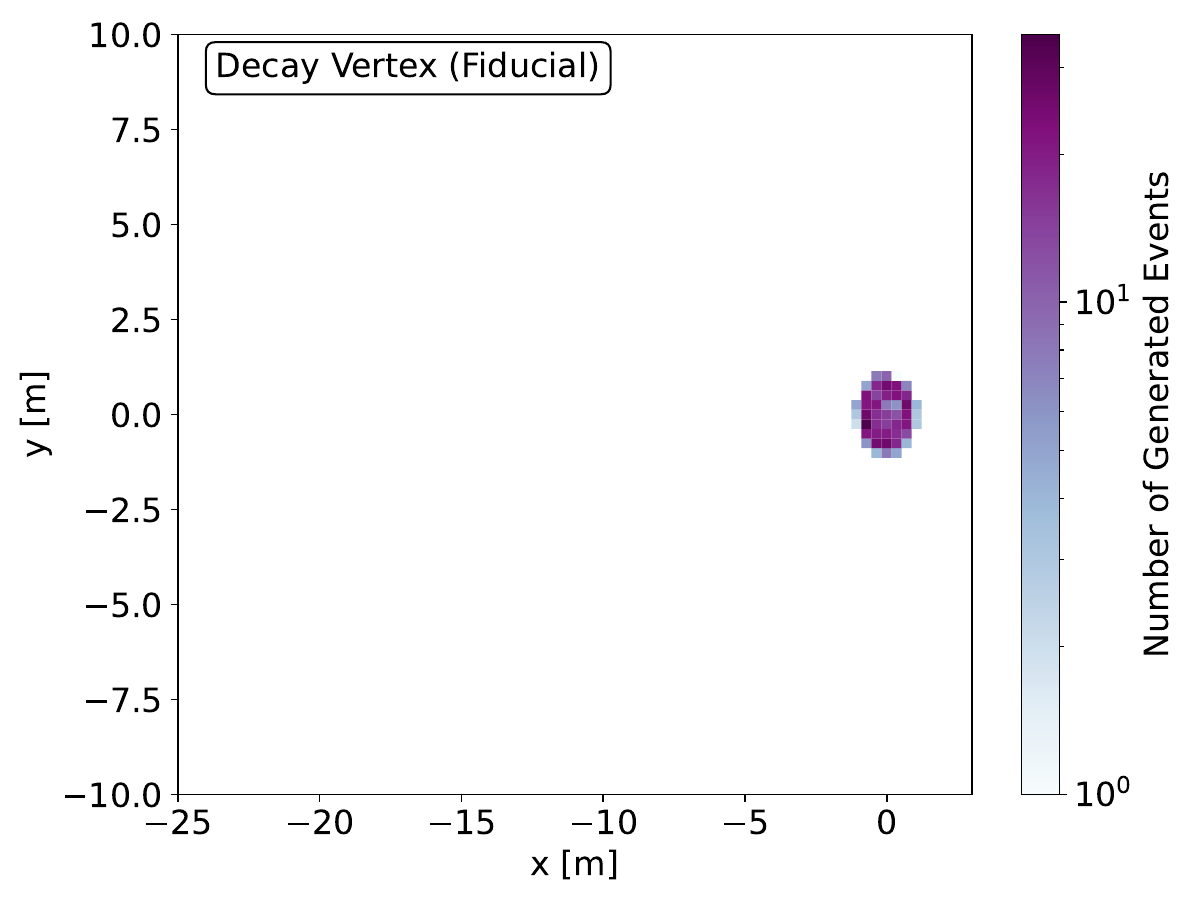}
    \caption{Sampled locations of the upscattering and decay interactions of the dipole-coupled HNL examples described in \cref{sec:dipole_examples}.
    The top, middle, and bottom rows correspond to MiniBooNE, \minerva, and CCM, respectively.
    The left and middle columns correspond to the upscattering and decay vertex, respectively.
    The rightmost column shows the decay vertices restricted to the fiducial volume of each experiment.}
    \label{fig:dipole_locations}
\end{figure*}

\begin{figure*}[h]
    \centering
    \includegraphics[width=0.9\textwidth]{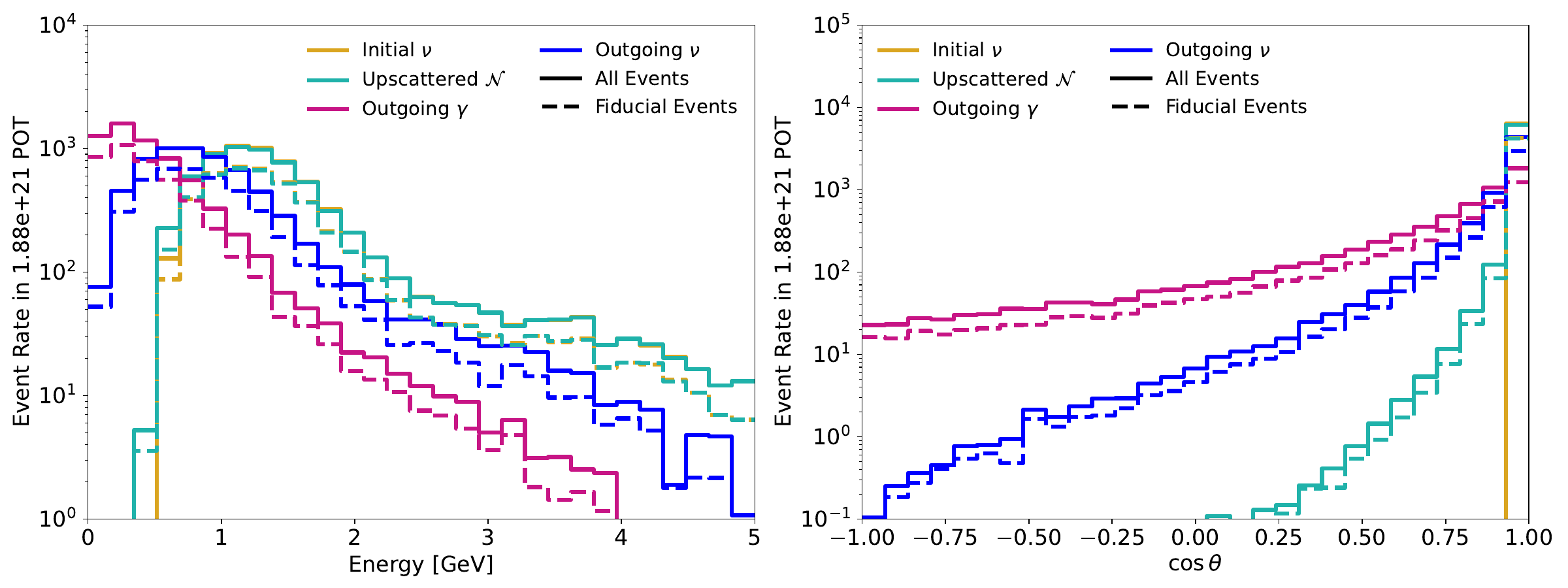}
    \\
    \includegraphics[width=0.9\textwidth]{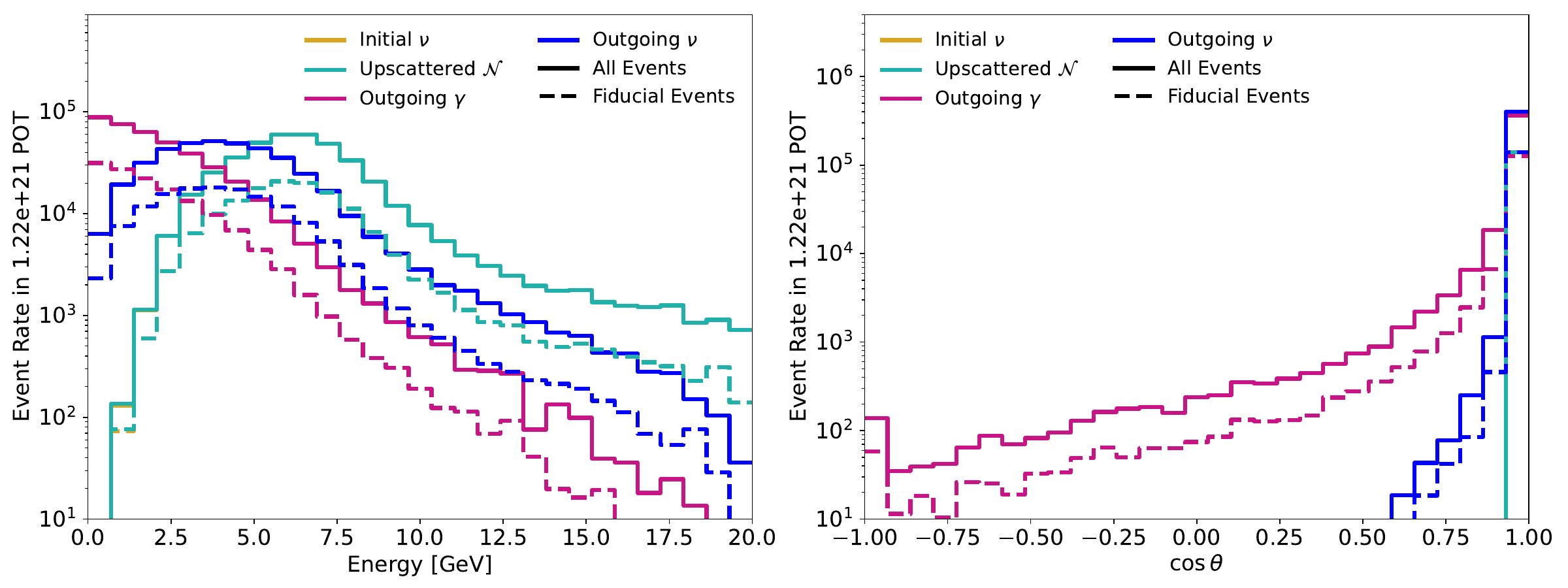}
    \caption{Physically-weighted energy (left) and angular (right) distributions of different particles in the $\nu A \to \mathcal{N} A$ and $\mathcal{N} \to \nu \gamma$ interactions of the dipole-coupled HNL model.
    The top and bottom rows correspond to the MiniBooNE and \minerva examples, respectively}
    \label{fig:dipole_kinematics}
\end{figure*}

The next example considers heavy neutral leptons (HNLs): right-handed partners of the left-handed Standard Model neutrinos with masses in the $\mathcal{O}({\rm MeV-GeV})$ regime.
These examples are intended to demonstrate \siren's ability to pair a variety of new physics models (via our \darknews interface) with detailed geometric descriptions of detectors.

We specifically explore HNLs which couple to the Standard Model neutrinos through an effective transition magnetic moment operator,
\begin{equation} {\label{eq:dipole_L}}
    \mathcal{L} \supset d_\alpha \overline{\mathcal{N}}_{R} \nu_{L \alpha} F_{\mu \nu} \sigma^{\mu \nu},
\end{equation}
where $d_{\alpha}$ is the transition magnetic moment, or dipole coupling, between the HNL $\mathcal{N}_R$ and the SM neutrino $\nu_{L \alpha}$, $F_{\mu \nu}$ is the field strength tensor of the photon, and $\sigma^{\mu \nu} = (i/2) [\gamma^\mu,\gamma^\nu]$.
This model has been explored extensively in the literature~\cite{Giunti:2008ve,deGouvea:2006hfo,Balantekin:2013sda,Vogel:1989iv,Kayser:1982br,Brdar:2021ysi,Magill:2018jla,Georgi:1990za,Babu:2021jnu} and has received particular attention as a target for experimental searches~\cite{Magill:2018jla,Brdar:2021ysi,Gninenko:1998nn,Gninenko:2009ks,Gninenko:2010pr,Gninenko:2012rw,Masip:2012ke,Coloma:2017ppo,Plestid:2020vqf,Schwetz:2020xra,Atkinson:2021rnp,Bolton:2021pey,Alvarez-Ruso:2021dna,Gustafson:2022rsz,Ovchynnikov:2022rqj,Zhang:2023nxy,Brdar:2023tmi}.
\Cref{eq:dipole_L} introduces several new interactions relevant for the production of these HNLs, including Dalitz-like neutral meson decays (e.g. $\pi^0 \to \gamma (\gamma^* \to \mathcal{N} \nu)$) and Primakoff upscattering ($\nu A \to \mathcal{N} A$).
The most commonly studied detection channel involves tagging the single photon from the radiative HNL decay ($\mathcal{N} \to \nu \gamma$).

Dipole-portal HNLs have been proposed as a potential explanation for the longstanding excess of electromagnetic shower-like events observed by the Mini Booster Neutrino Experiment (MiniBooNE)~\cite{Gninenko:2009ks,McKeen:2010rx,Gninenko:2010pr,Dib:2011jh,Gninenko:2012rw,Masip:2012ke,Radionov:2013mca,Ballett:2016opr,Magill:2018jla,Balantekin:2018ukw,Balaji:2019fxd,Balaji:2020oig,Fischer:2019fbw,Alvarez-Ruso:2021dna,Vergani:2021tgc}.
In particular, Ref.~\cite{Kamp:2022bpt} used an earlier iteration of \siren to simulate these interactions in MiniBooNE and thus determine the preferred region in $m_\mathcal{N}-d_{\mu}$ parameter space to explain the MiniBooNE excess.
This study also computed constraints on this model using elastic scattering measurements~\cite{MINERvA:2015nqi,Valencia:2019mkf,MINERvA:2022vmb} from the Main Injector Neutrino ExpeRiment to study $\nu-A$ interactions (\minerva), which required detailed geometric modeling of the complex subcomponents of the \minerva detector~\cite{MINERvA:2013zvz}.
Further improvements were made to this earlier iteration of \siren in order to study dipole-portal HNLs within the Coherent CAPTAIN-Mills (CCM) experiment~\cite{Kamp:2023hai}.
CCM uses a light-based liquid argon detector to search for particles produced in the Lujan proton beam dump source of the Los Alamos Neutron Science Center (LANSCE)The CCM detector operates at the Lujan beam dump facility of the Los Alamos Neutron Science Center (LANSCE)~\cite{NELSON2012172,LISOWSKI2006910}.
We are interested in the monoenergetic 30\;MeV muon neutrinos produced by pion decay at rest~\cite{CCM:2021leg}.
In the dipole-portal model, these muon neutrinos can produce HNLs via upscattering in the tungsten beam dump target or the surrounding shielding, which can then decay inside the CCM detector.
Estimating the event rate from this process thus requires a detailed simulation of the CCM detector hall, which is possible through the flexible geometry interface of \siren.

The three examples are set up similarly--they all simulate the production of Dirac HNLs via Primakoff upscattering in and around the detector volume as well as the single photon decay of HNLs within the detector volume.
The cross section and decay width calculations are handled by the \darknews interface of \siren described in \cref{sec:dark_news}.

The MiniBooNE and \minerva examples are based on the study presented in Ref.~\cite{Kamp:2022bpt}.
For MiniBooNE, we consider HNL production in the bedrock along the Booster Neutrino Beam (BNB) and within MiniBooNE itself.
The MiniBooNE detector model is described in detail in \cref{sec:provided_detectors}.
We sample the initial energy of the muon neutrino using the tabulated MiniBooNE BNB flux prediction~\cite{MiniBooNE:2008hfu}.
For \minerva, we consider HNL production in the bedrock along the Neutrino Main Injector (NuMI) beamline and within various detector subcomponents.
Many subcomponents are considered, including the veto wall, nuclear target region, active tracker region, and electromagnetic calorimeter.
The last three of these are implemented using extruded polygons, allowing us to accurately model the hexagonal prism structure of the \minerva detector.
We also correctly model the structure of the nuclear target layers, in which different nuclear targets comprise different subsections of the hexagonal prism as described in Ref.~\cite{MINERvA:2013zvz}.
More details about the \minerva detector model are provided in \cref{sec:provided_detectors}.
Neutrino energies are sampled using the medium-energy NuMI flux digitized from Ref.~\cite{AliagaSoplin:2016shs}.
In both the MiniBooNE and \minerva examples, we sample upscattering locations using ranged injection with respect to the decay length of the HNL.
Once a neutrino path is determined, the actual interaction position along that path is sampled according to the interaction length, which depends non-trivially on the traversed materials (in contrast to the DIS case, where we can sample according to column depth).

The CCM example is based on the study presented in Chapter~7 of Ref.~\cite{Kamp:2023hai}.
Here, we consider HNL production in any of the materials between the Lujan beam dump target and the CCM detector, described in detail in \cref{sec:provided_detectors}.
We generate 30\;MeV muon neutrinos emitted isotropically from the tungsten target to simulate pion decay-at-rest.
The total neutrino flux is taken from Ref.~\cite{CCM:2021leg}.
We use a point source position distribution, which samples the upscattering location according to the interaction length from the center of the tungsten target along the neutrino direction out to a maximum distance of 25\;m, i.e. beyond the CCM detector.

In \cref{fig:dipole_locations}, we show distributions of the generated upscattering and decay locations for dipole-portal interactions in MiniBooNE, \minerva, and CCM.
The upscattering locations reveal the different subcomponents of the detector environment.
In the MiniBooNE and \minerva cases the HNLs are relatively short-lived, such that most upscattering interactions are sampled within or near the detector.
In the \minerva case specifically, one can see the upstream veto wall as well as the higher-$Z$ nuclear targets in the forward part of the detector.
Ref.~\cite{Kamp:2022bpt} discusses in more detail the imprint of the detailed geometric configuration of in \minerva in our \siren simulation.
HNLs are longer-lived in the CCM case; therefore the entire detector hall becomes relevant to the calculation.
Most HNLs are generated within the TMRS, though the additional downstream shielding and the detector itself are also visible.
The decay locations are a convolution of the upscattering locations and the decay length of the HNL.
Additionally, if the direction of the HNL intersects the fiducial volume, the decay is required to occur within the fiducial volume to increase computational efficiency.
Thus one can see the imprint of each experiment's fiducial volume in the decay location distribution.
This is made clear by the third column, which shows the decay location distributions restricted to the fiducial volume.

In \cref{fig:dipole_kinematics}, we show the physically-weighted energy and angular distribution of each particle in the dipole interaction chain for the MiniBooNE and \minerva examples.
The angle here is computed with respect to the beam axis.
We also show the same distributions restricted to events for which the decay occurs within the fiducial volume.
The overall normalization is scaled to match the total collected protons-on-target for the forward horn current BNB mode and reverse horn current medium energy NuMI mode for MiniBooNE and \minerva, respectively. 
To first order, the neutrino energy distributions reflect the BNB or NuMI flux convoluted with the energy dependence of the upscattering cross section.
The angular distributions reveal that despite the preference for low momentum transfer (and thus forward-going HNLs) in the Primakoff upscattering process, the final state photons develop a non-negligible large-angle component.
This effect is more prominent in MiniBooNE than in \minerva due to the lower typical energies of the BNB compared to NuMI.
Furthermore, one can see a difference in the energy and angular distributions of the outgoing photon and neutrino, which is a direct consequence of the Dirac nature of the HNL~\cite{Balantekin:2018ukw,Alvarez-Ruso:2021dna}.
Accurate estimations of the kinematic distributions of observable final state particles are necessary for robust fits to experimental data in the MiniBooNE case and accurate application of kinematic cuts in the \minerva case~\cite{Kamp:2022bpt}.

%% file: sections/07_conclusion.tex
\section{Conclusion and Future Directions}
\label{sec:conclusion}

This article has presented the \siren software package, an open-source toolkit enabling the efficient simulation of rare neutrino interactions in complex detector geometries.
The extensibility of \siren makes it straightforward for the user to study a variety of interaction models, including potential BSM interactions, within a variety of detector geometries.
The injection methodologies supported by \siren allow the user to sample from biased distributions that optimize simulation efficiency for the interaction model and detector geometry under consideration.
Our comprehensive weighting interface removes the effect of these biased generation distributions.
By saving information about the distributions from which events were sampled during generation, \siren can reweight these events to any desired physical interaction model, detector geometry, or distribution related to the primary particle (e.g., the energy distribution of the initial neutrino).
This reweighting scheme is essential when \siren is used to feed more detailed detector response simulations.

We have demonstrated the potential use cases of \siren through two examples, the first exploring the detection of muons from $\nu_\mu$ DIS and the second exploring the detection of photons from the decay of dipole-coupled HNLs in accelerator neutrino experiments.
The latter of these leverages the \cname{DarkNews} interface of \siren, which pairs the extensive suite of HNL-based interaction models supported by \cname{DarkNews} with the flexible geometry description and efficient reweightability provided by \siren.

The authors envision several updates to this initial version of \siren that will extend its capabilities and use cases.
First, we plan to pair \siren with other neutrino event generators focused on detailed neutrino cross section calculations, including \cname{GENIE}~\cite{Andreopoulos:2015wxa}, \cname{ACHILLES}~\cite{Isaacson:2022cwh}, and \cname{MARLEY}~\cite{Gardiner:2021qfr}.
These interfaces would be modeled off of the existing \cname{DarkNews} interface and would significantly enlarge the set of SM and BSM neutrino interaction models supported within \siren.
We are also interested in extending the BSM models supported by \siren at the time of writing into the DIS regime.
This would allow \siren to explore exotic signatures of new physics in the high energy atmospheric and astrophysical neutrino flux at neutrino telescopes such as IceCube, KM3NeT~\cite{KM3Net:2016zxf}, and Baikal-GVD~\cite{Baikal-GVD:2018isr}.

Future versions of \siren will also include more detector geometry configuration files immediately available to users.
These include (but are not limited to) ND280~\cite{T2K:2019bbb}, the Short Baseline Neutrino program (MicroBooNE, SBND, and ICARUS)~\cite{MicroBooNE:2015bmn}, and the DUNE near detector~\cite{DUNE:2021tad}.
We also plan to support the construction of detector models through the GDML interface~\cite{Chytracek:2006be}.
Finally, we will make the \siren output compatible with the \cname{Prometheus} open-source neutrino telescope simulation~\cite{Lazar:2023rol}, which can translate events injected by \siren into photons observed by current and next-generation neutrino telescopes.

%% file: sections/08_acknowledge.tex
\section*{Acknowledgements}
\begin{CJK*}{UTF8}{gbsn}

The authors thank Jackapan Pairin for the artistic rendering in \cref{fig:siren_injection_diagram}, Matheus Hostert for discussions regarding the \darknews interface, Carlos Arg\"uelles for suggestions on the manuscript, Janet Conrad for presenting the problems that inspired this work, and the authors of \prevleptoninjector for their prior work on these issues.
AS is supported by the U.S. Department of Energy through the Los Alamos National Laboratory.
Los Alamos National Laboratory is operated by Triad National Security, LLC, for the National Nuclear Security Administration of U.S. Department of Energy (Contract No. \seqsplit{89233218CNA000001}).
NK was supported by the National Science Foundation (NSF) CAREER Award 2239795 and the David and Lucile Packard Foundation.
AYW was supported by the Harvard Physics Department Purcell Fellowship and the Natural Sciences and Engineering Research Council of Canada (NSERC), funding reference number PGSD-577971-2023. 

\end{CJK*}

%% file: appendices/a_computational.tex
\section{Computational Efficiency}
\label{app:comp_efficiency}

\begin{figure}
    \centering
    \includegraphics[width=\linewidth]{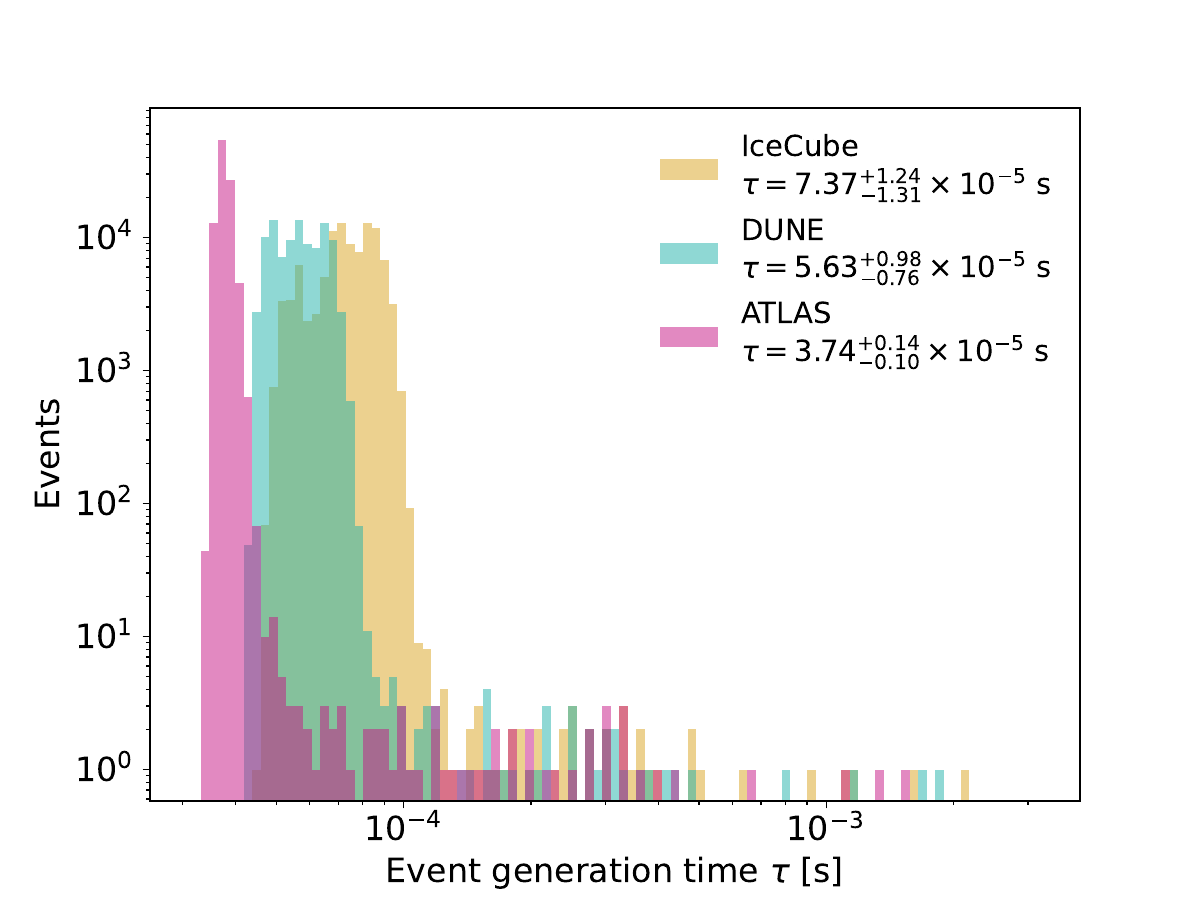}
    \includegraphics[width=\linewidth]{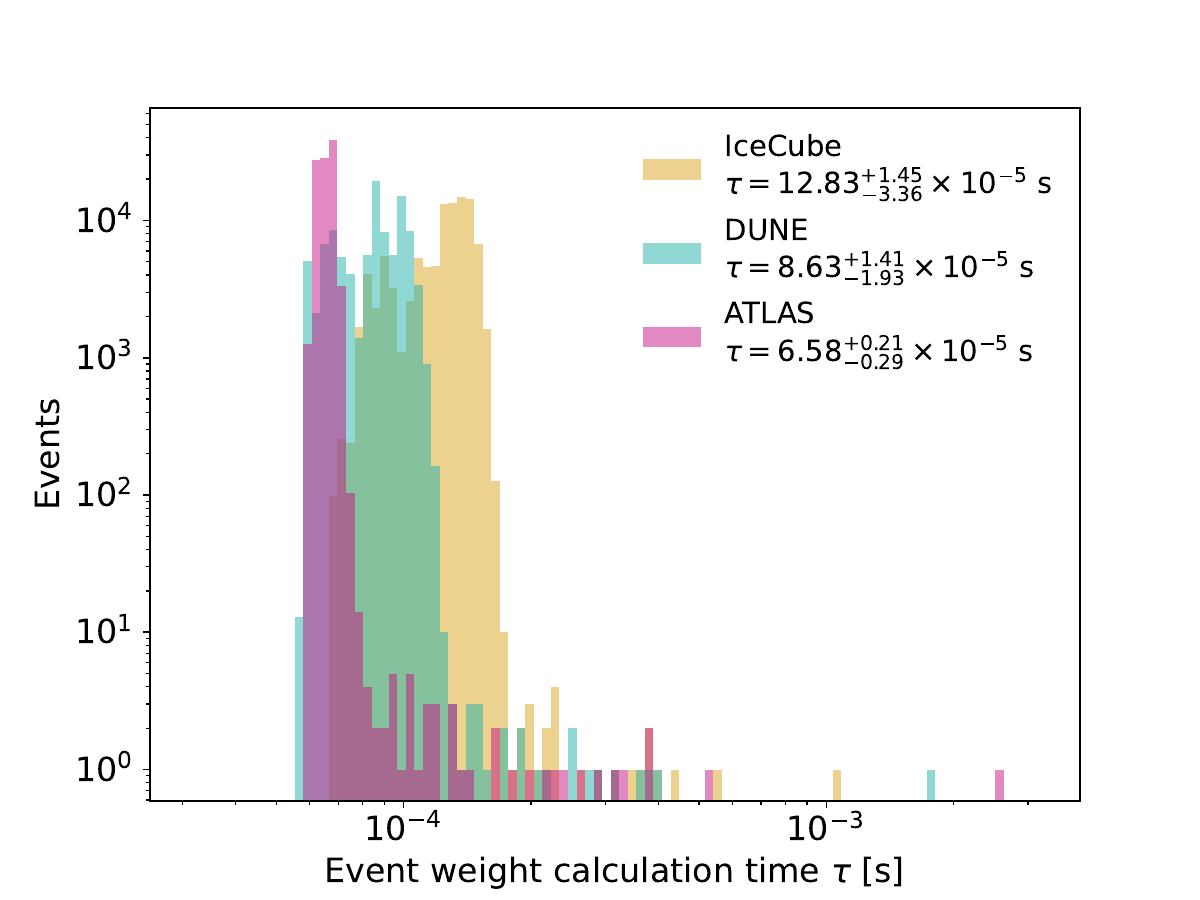}
    \caption{The event generation time (top) and weight calculation (bottom) distributions for the DIS examples presented in \cref{sec:DIS_examples}. The values in the legend reflect the median and $\pm 1\sigma$ width of each distribution.}
    \label{fig:DIS_timing}
\end{figure}

\begin{figure}
    \centering
    \includegraphics[width=\linewidth]{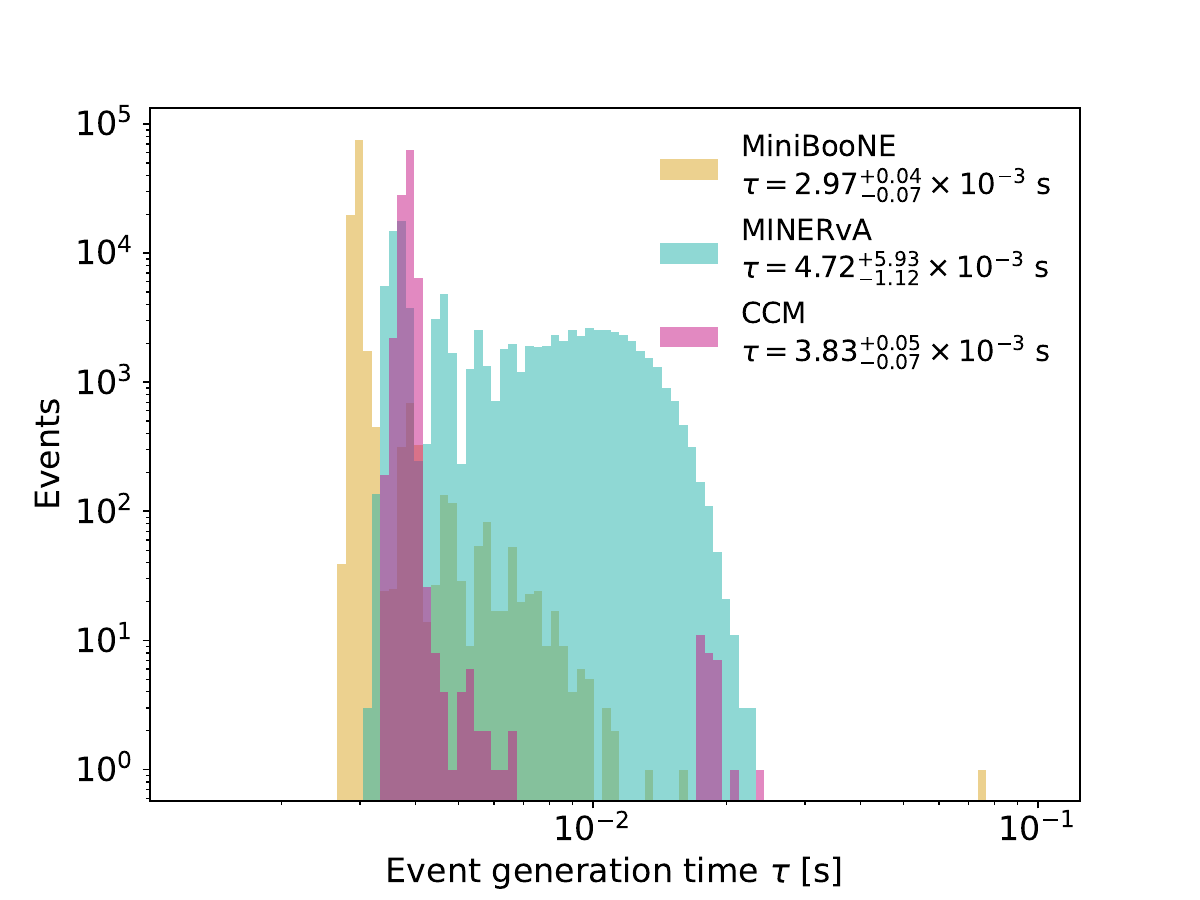}
    \includegraphics[width=\linewidth]{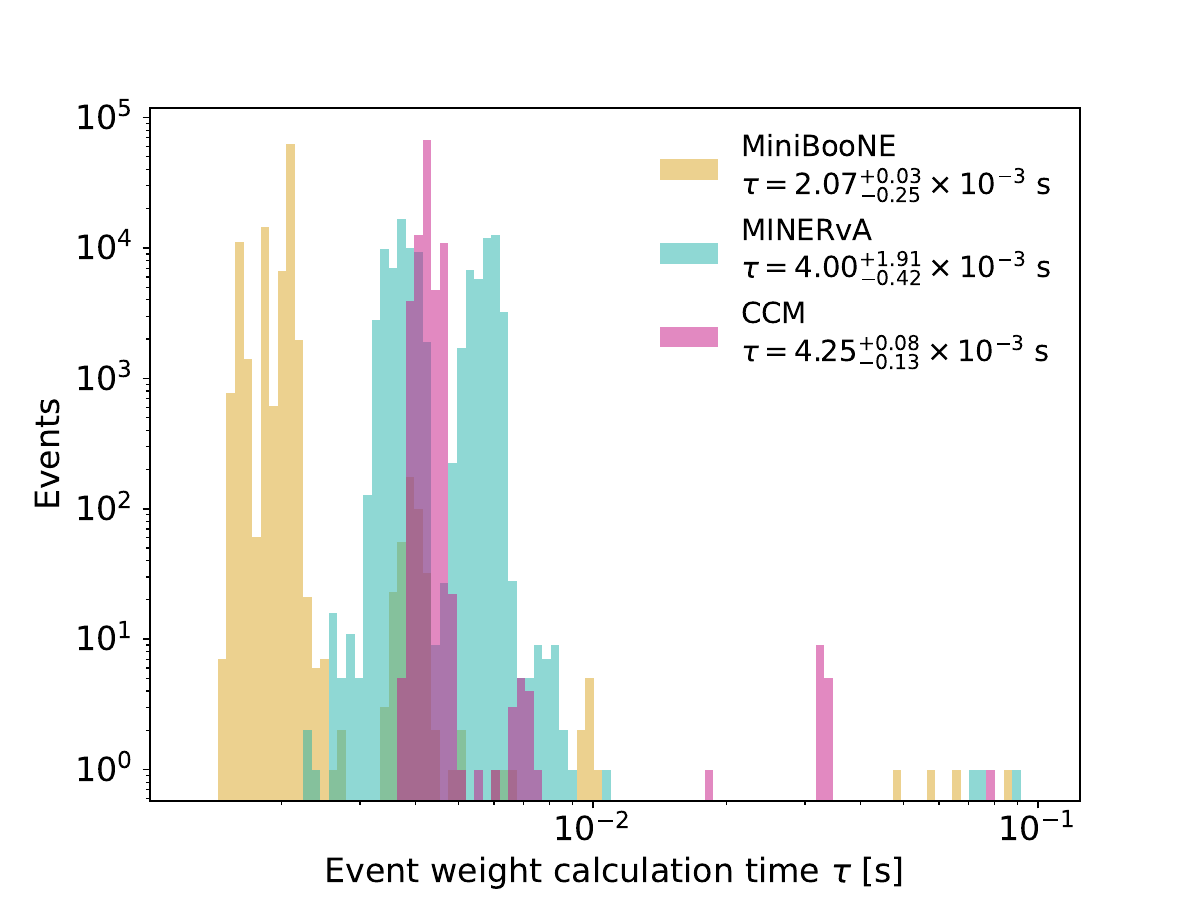}
    \caption{The event generation time (top) and weight calculation (bottom) distributions for the HNL examples presented in \cref{sec:dipole_examples}. The values in the legend reflect the median and $\pm 1\sigma$ width of each distribution.}
    \label{fig:dipole_timing}
\end{figure}

This appendix provides distributions of event generation times and weight calculation times for the examples shown in \cref{sec:examples}, as demonstrated in \cref{fig:DIS_timing,fig:dipole_timing}.
These distributions were used to generate the values in \cref{tab:computation_efficiency}.
\Cref{fig:DIS_timing,fig:dipole_timing} correspond to the DIS examples in \cref{sec:DIS_examples} and the HNL examples in \cref{sec:dipole_examples}, respectively.

%% file: appendices/b_code_examples.tex
\section{Code Examples}
\label{app:code_examples}

This appendix includes two example \cname{Python} scripts for event generation and weighting in \siren.
These can be found in the \cname{resources/Examples/} directory in the repository along with similar scripts for the other examples presented in \cref{sec:examples}.

\subsection{$\nu_\mu$ DIS in IceCube}

\begin{lstlisting}
import os

import siren
from siren.LIController import LIController

# Number of events to inject
events_to_inject = int(1e5)

# Expeirment to run
experiment = "IceCube"

# Define the controller
controller = LIController(events_to_inject, experiment)

# Particle to inject
primary_type = siren.dataclasses.Particle.ParticleType.NuMu

cross_section_model = "CSMSDISSplines"

xsfiledir = siren.utilities.get_cross_section_model_path(cross_section_model)

# Cross Section Model
target_type = siren.dataclasses.Particle.ParticleType.Nucleon

DIS_xs = siren.interactions.DISFromSpline(
    os.path.join(xsfiledir, "dsdxdy_nu_CC_iso.fits"),
    os.path.join(xsfiledir, "sigma_nu_CC_iso.fits"),
    [primary_type],
    [target_type], "m"
)

primary_xs = siren.interactions.InteractionCollection(primary_type, [DIS_xs])
controller.SetInteractions(primary_xs)

# Primary distributions
primary_injection_distributions = {}
primary_physical_distributions = {}

mass_dist = siren.distributions.PrimaryMass(0)
primary_injection_distributions["mass"] = mass_dist
primary_physical_distributions["mass"] = mass_dist

# energy distribution
edist = siren.distributions.PowerLaw(2, 1e3, 1e6)
primary_injection_distributions["energy"] = edist
primary_physical_distributions["energy"] = edist

# direction distribution
direction_distribution = siren.distributions.IsotropicDirection()
primary_injection_distributions["direction"] = direction_distribution
primary_physical_distributions["direction"] = direction_distribution

# position distribution
muon_range_func = siren.distributions.LeptonDepthFunction()
position_distribution = siren.distributions.ColumnDepthPositionDistribution(
    600, 600.0, muon_range_func, set(controller.GetDetectorModelTargets()[0])
)
primary_injection_distributions["position"] = position_distribution

# SetProcesses
controller.SetProcesses(
    primary_type, primary_injection_distributions, primary_physical_distributions
)

controller.Initialize()

events = controller.GenerateEvents()

os.makedirs("output", exist_ok=True)

controller.SaveEvents("output/IceCube_DIS")
\end{lstlisting}

\subsection{Dipole-portal HNLs in MiniBooNE}

\begin{lstlisting}
import os

import siren
from siren.SIREN_Controller import SIREN_Controller

# Define a DarkNews model
model_kwargs = {
    "m4": 0.47,  # 0.140,
    "mu_tr_mu4": 2.50e-6,  # 1e-6, # GeV^-1
    "UD4": 0,
    "Umu4": 0,
    "epsilon": 0.0,
    "gD": 0.0,
    "decay_product": "photon",
    "noHC": True,
    "HNLtype": "dirac",
}

# Number of events to inject
events_to_inject = 100000

# Expeirment to run
experiment = "MiniBooNE"

# Define the controller
controller = SIREN_Controller(events_to_inject, experiment)

# Particle to inject
primary_type = siren.dataclasses.Particle.ParticleType.NuMu

xs_path = siren.utilities.get_cross_section_model_path(f"DarkNewsTables-v{siren.utilities.darknews_version()}", must_exist=False)
# Define DarkNews Model
table_dir = os.path.join(
    xs_path,
    "Dipole_M%2.2e_mu%2.2e" % (model_kwargs["m4"], model_kwargs["mu_tr_mu4"]),
)
controller.InputDarkNewsModel(primary_type, table_dir, **model_kwargs)

# Primary distributions
primary_injection_distributions = {}
primary_physical_distributions = {}

# energy distribution
flux_file = siren.utilities.get_tabulated_flux_file("BNB","FHC_numu")
edist = siren.distributions.TabulatedFluxDistribution(flux_file, True)
edist_gen = siren.distributions.TabulatedFluxDistribution(
    model_kwargs["m4"], 10, flux_file, False
)
primary_injection_distributions["energy"] = edist_gen
primary_physical_distributions["energy"] = edist

# direction distribution
direction_distribution = siren.distributions.FixedDirection(siren.math.Vector3D(0, 0, 1.0))
primary_injection_distributions["direction"] = direction_distribution
primary_physical_distributions["direction"] = direction_distribution

# position distribution
decay_range_func = siren.distributions.DecayRangeFunction(
    model_kwargs["m4"], controller.DN_min_decay_width, 3, 541
)
position_distribution = siren.distributions.RangePositionDistribution(
    6.2, 6.2, decay_range_func, set(controller.GetDetectorModelTargets()[0])
)
primary_injection_distributions["position"] = position_distribution

# SetProcesses
controller.SetProcesses(
    primary_type, primary_injection_distributions, primary_physical_distributions
)

controller.Initialize()

def stop(datum, i):
    secondary_type = datum.record.signature.secondary_types[i]
    return secondary_type != siren.dataclasses.Particle.ParticleType.N4

controller.injector.SetStoppingCondition(stop)

events = controller.GenerateEvents(fill_tables_at_exit=False)

os.makedirs("output", exist_ok=True)

controller.SaveEvents(
    "output/MiniBooNE_Dipole_M%2.2e_mu%2.2e_example"
    % (model_kwargs["m4"], model_kwargs["mu_tr_mu4"]),
    fill_tables_at_exit=False
)
\end{lstlisting}

%% file: appendices/c_lepton_injector_validation.tex
\section{Validation Against \prevleptoninjector}
\label{app:LI_validation}

\begin{figure}
    \centering
    \includegraphics[width=0.5\textwidth]{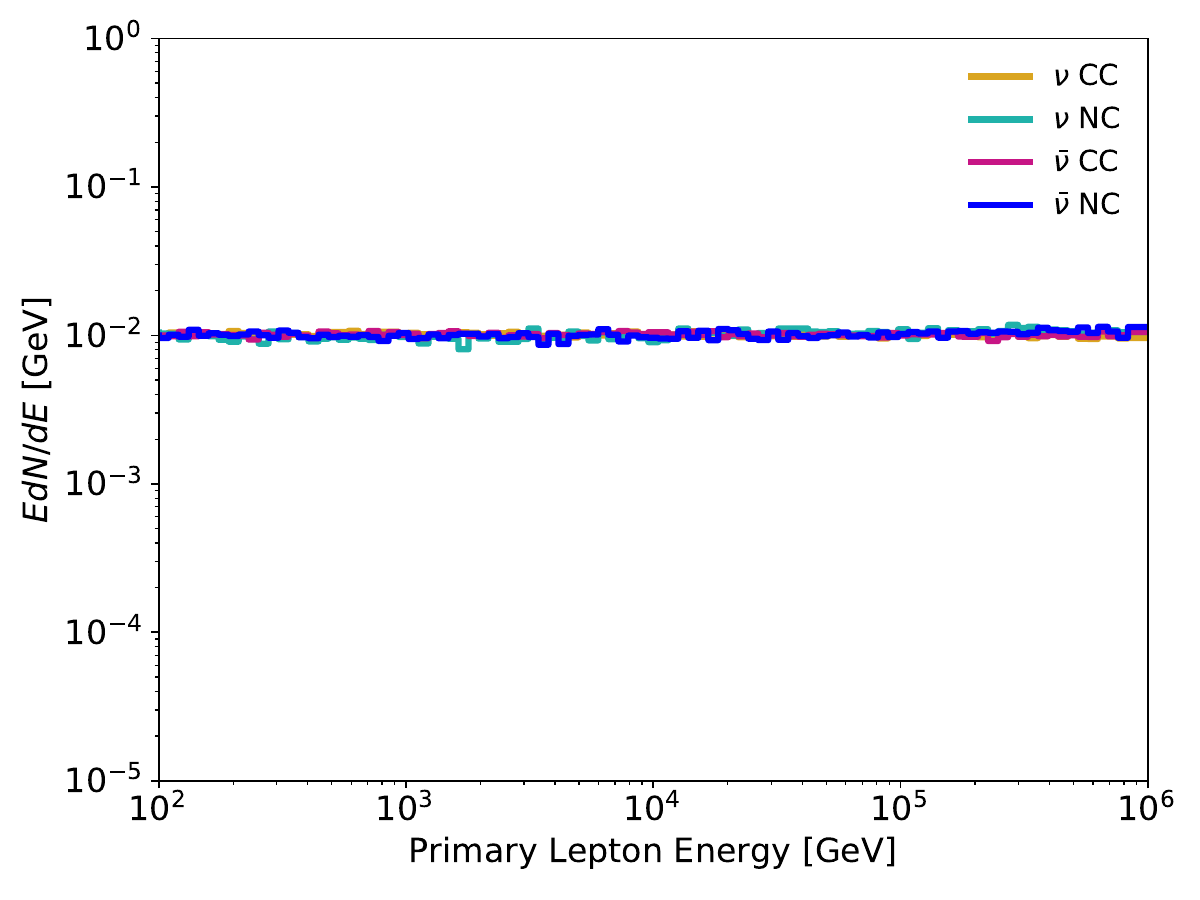}
    \includegraphics[width=0.5\textwidth]{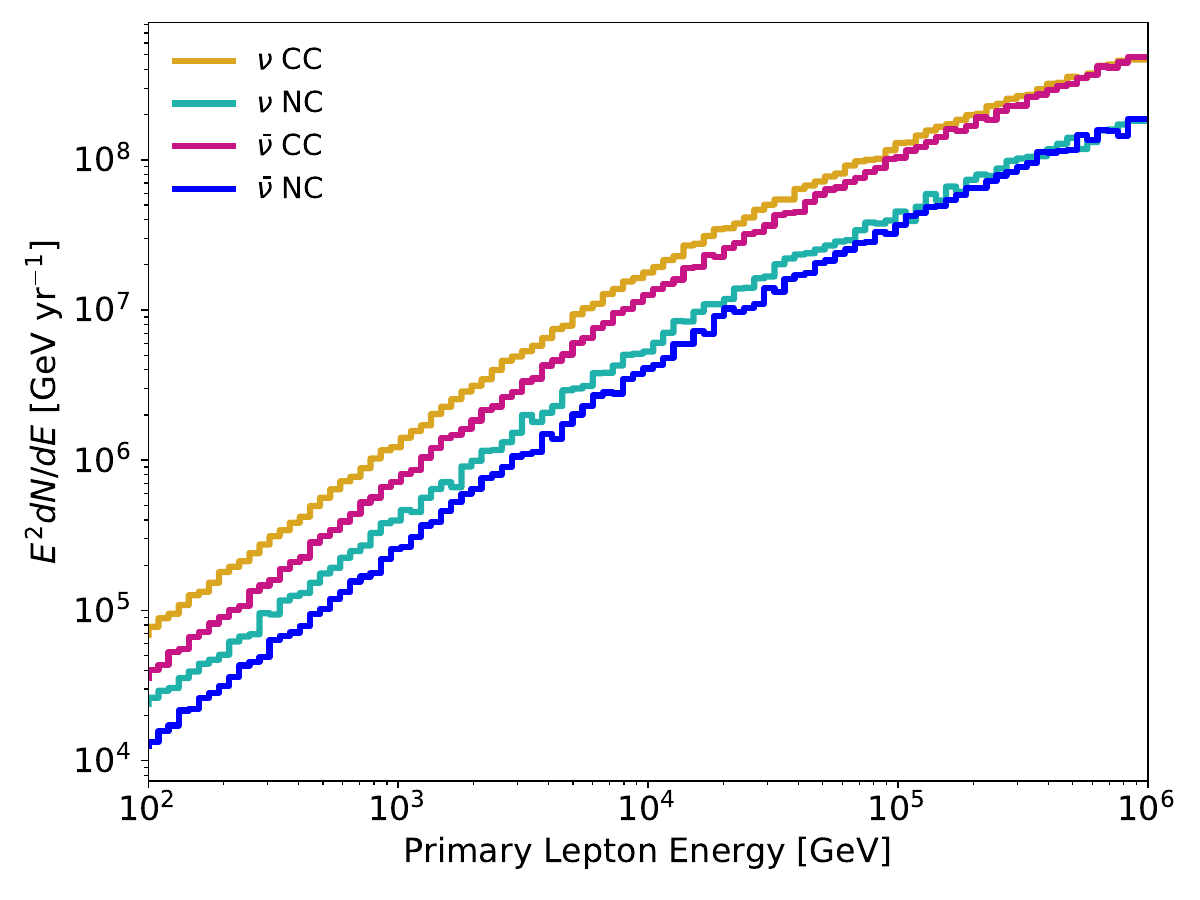}
    \includegraphics[width=0.5\textwidth]{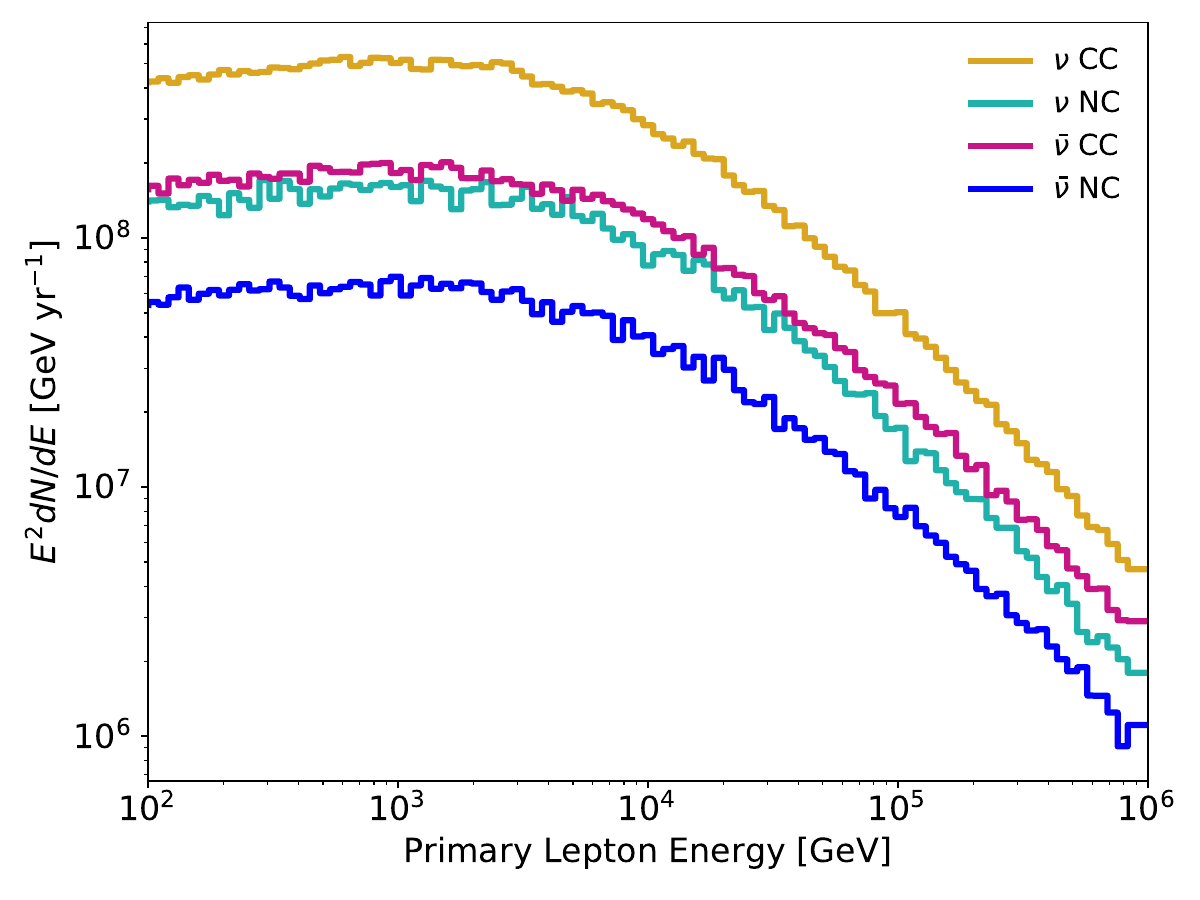}
    \caption{Following Fig.~3.1 of Ref.~\cite{IceCube:2020tcq}, this figure shows distributions of the primary neutrino energy for a \siren sample of all-flavor $\nu$ DIS events in and around IceCube with primary neutrino energies sampled from an $E^{-1}$ power law distribution.
    The top panel shows the unweighted distribution for $\nu$ and $\bar{\nu}$ CC and NC events.
    The middle panel shows the event rate distribution reweighted to an astrophysical $E^{-2}$ flux and the CSMS $\nu$ DIS cross section calculation~\cite{Cooper-Sarkar:2011jtt}.
    The bottom panel shows the event rate distribution reweighted to the atmospheric flux calculation of Ref.~\cite{Honda:2006qj}, again using the CSMS cross section calculation}
    \label{fig:LI_validation}
\end{figure}

This appendix discusses the backward compatibility of \siren.
This is reflected by \siren's ability to perform the use cases of its predecessor, \prevleptoninjector; namely, the injection of $\nu$ DIS interactions in the IceCube detector.
To do this, we follow the reweighting exercise depicted in Fig.~3.1 of Ref.~\cite{IceCube:2020tcq}.
We begin by injecting all-flavor neutrinos according to an $E^{-1}$ power law distribution.
We then reweight this sample to compute the physical event rate assuming an $E^{-2}$ astrophysical flux and using the CSMS calculation of the $\nu$ DIS cross section~\cite{Cooper-Sarkar:2011jtt}.
Following Ref.~\cite{IceCube:2020tcq}, we also reweight these events to the atmospheric neutrino flux calculation from Ref.~\cite{Honda:2006qj}, again using the CSMS cross section calculation.
\Cref{fig:LI_validation} shows the generation-level (i.e. unweighted) primary neutrino energy distribution as well as the physical neutrino energy distributions for the astrophysical and atmospheric flux cases.
These distributions are consistent with those presented in Ref.~\cite{IceCube:2020tcq}.

